\documentclass[sigconf,screen,nonacm]{acmart}

\newif\ifeditversion
\editversionfalse

\ifeditversion
  \colorlet{revisioncolor}{blue}
\else
  \colorlet{revisioncolor}{black}
\fi

\newcommand{\mycolor}[1]{\textcolor{revisioncolor}{#1}}

\newif\ifrevisiontwo
\revisiontwofalse
\ifrevisiontwo
  \colorlet{revisiontwocolor}{red}
  \newcommand{\recolor}[1]{\textcolor{revisiontwocolor}{#1}}
\else
  \newcommand{\recolor}[1]{#1}
\fi

\usepackage{cleveref}
\usepackage{algorithm}
\usepackage{algorithmic}
\usepackage{graphicx}
\usepackage{multirow}
\usepackage{makecell}

\AtBeginDocument{%
  }

\begin{document}

\title{Toward Natural and Companionable Virtual Agents via Cross-Temporal Emotional Modeling}

\author{Feier Qin}
\affiliation{%
  \institution{Communication University of China}
  \city{BeiJing}
  \country{China}
}
\email{feiqin2323@gmail.com}

\author{Xiao Li}
\affiliation{%
  \institution{Microsoft Research Asia}
  \city{BeiJing}
  \country{China}
}
\email{xili11@microsoft.com}

\author{Yi Zheng}
\affiliation{%
  \institution{Communication University of China}
  \city{BeiJing}
  \country{China}
}
\email{zhengyi7592@cuc.edu.cn}

\author{Haibin Huang}
\affiliation{%
  \institution{Institute of Artificial Intelligence, China Telecom}
  \city{BeiJing}
  \country{China}
}
\email{jackiehuanghaibin@gmail.com}

\author{Hanyao Wang}
\affiliation{%
  \institution{Communication University of China}
  \city{BeiJing}
  \country{China}
}
\email{1944597678@qq.com}

\author{Xiaoyu Wang}
\affiliation{%
  \institution{Communication University of China}
  \city{BeiJing}
  \country{China}
}
\email{2718386675@qq.com}

\author{Yan Lu}
\affiliation{%
  \institution{Microsoft Research Asia}
  \city{BeiJing}
  \country{China}
}
\email{yanlu@microsoft.com}

\author{Yuan Zhang}
\affiliation{%
  \institution{Communication University of China}
  \city{BeiJing}
  \country{China}
}
\email{yzhang@cuc.edu.cn}

\renewcommand{\shortauthors}{Qin et al.}

\begin{abstract}
Recent advances in foundation models have enabled conversational agents that aim for sustained companionship rather than mere task completion.
Yet most still remain unable to support natural, long-term companion-like interactions, resulting in experiences that feel episodic and inauthentic. We argue that current agents overlooked cross-temporal modeling of agents’ social behaviors and internal emotions: generated behaviors rarely influence an agent’s emotional state, and emotional states seldom shape subsequent behaviors.
We present Cross-Temporal Emotion Modeling (CTEM), a framework that links long-term behavioral history to moment-to-moment emotional expression. CTEM establishes a closed loop where past experiences update an evolving emotional state; this state conditions immediate interactions; and user feedback continually revises both memory and emotional state, enabling reflection and anticipation.
We instantiate CTEM as \textit{Auri}, a companion agent on an instant-messaging platform, and report a 21-day in-the-wild study showing that CTEM shows improvements in perceived naturalness, coherence, and emotional harmony.
\end{abstract}



\begin{teaserfigure}
  \includegraphics[width=\textwidth]{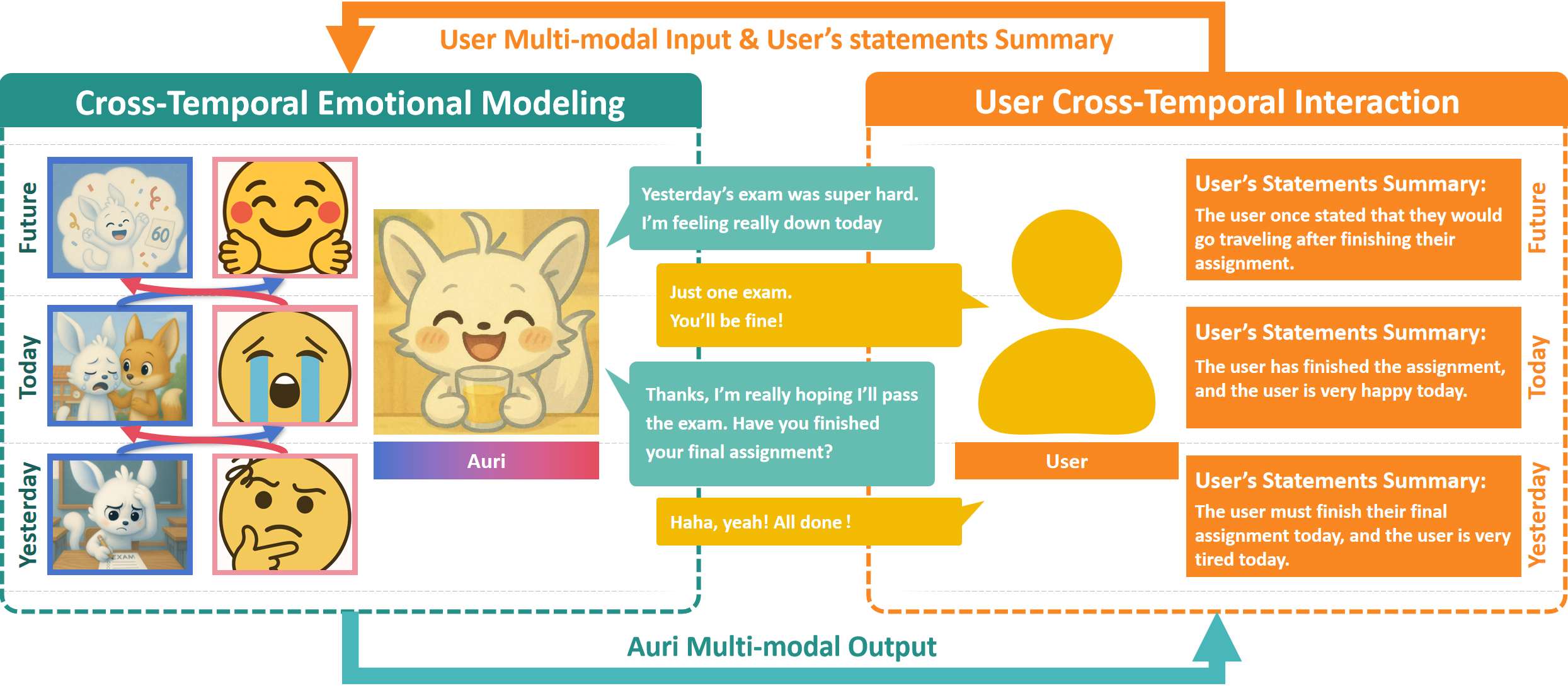}
  \caption{We present \textit{Auri}, a lightweight companion agent designed to foster long-term emotional connections through cross-temporal emotional modeling. 
          \textit{Auri} is able to deliver contextually coherent and emotionally resonant interactions over time.}
  \Description{}
  \label{fig:teaser}
\end{teaserfigure}

\maketitle

\section{Introduction}
Virtual agents are increasingly envisioned not merely as task assistants, but as companions in education, healthcare, and everyday social interaction~\cite{park2023generative_agents,maples2024loneliness}. 
Unlike task-oriented dialogue systems, companion agents must sustain relationships that feel coherent and emotionally engaging over time.
To achieve this, a companion agent should have its own behavior model that builds on past interactions to sustain coherent actions while remaining emotionally attuned in each moment.
Psychological and neuroscientific research~\cite{frijda1986emotions,mcclay2023dynamic, Ochsner2007, maslow1943theory, wigfield2000expectancy} has revealed that emotions systematically shape humans' habits and behaviors (and vice versa) over time; 
therefore, companion agents for human–agent interaction should likewise model the coupling between evolving affect and behavioral trajectories.

However, the behavior model for existing agentic approaches remains incomplete, as they often treat behavior and affect in isolation.
For example, recent agentic social simulation approaches~\cite{park2023generative_agents} model how agents remember and plan their behaviors over time, achieving continuity in their actions similar to humans. However, these approaches rarely incorporate an internal affective model, and thus cannot capture how an agent’s emotions evolve across interactions or how these emotions subsequently guide future behaviors.
Many conversational agent systems~\cite{pataranutaporn2024future,das2025ai} typically rely on predefined personalities or emotional profiles, enabling vivid and empathic responses at moments. However, these personalities remain static and disconnected from the agent’s lived experiences; such systems do not account for how past interactions progressively shape the agent’s internal affective state or how this evolving state influences subsequent behavior.
Some methods also attempt to track evolving emotions~\cite{croissant2024chain_of_emotion}, but they remain limited to local dialogue appraisals and do not model emotions as jointly shaped by the agent’s own behaviors and user feedback over time.
\mycolor{These limitations highlight the need for computational models that explicitly couple long-term behavioral accumulation with immediate affective expression in a closed loop. 
Establishing this coupling is essential for fostering and understanding the \textit{perceived coherence} (the agent’s consistency in behaviors, personality, and characteristics across time) and \textit{emotional harmony} (the user’s sense of being emotionally supported and appropriately responded to in their cross-temporal contextual cues) of companionship.}

In this paper, we take a step toward addressing this gap by proposing \textbf{Cross-Temporal Emotional Modeling (CTEM)}, a framework that formalizes how agents’ behavioral histories and emotional states mutually influence one another across time.
CTEM comprises three mechanisms inspired by psychological theories of emotion and behavior:
(1) a psychological-grounded behavior pool with dynamic generation and integration;
(2) internal states of agent emotion, personalities and history memories that evolve over time; and 
(3) a multi-modal adaptive interaction scheme with human users, leveraging the latest foundation models~\cite{hurst2024gpt}.

We instantiate the CTEM framework as \textbf{\textit{Auri}}, a lightweight companion agent deployed on an instant messaging platform. For evaluation, we conducted a 21-day in-the-wild user study with \mycolor{96} participants aged 18--26, collecting data through computational logs, surveys, interviews, and discussions.
Our study addresses the following research questions:
\begin{itemize}
    \item \textbf{RQ1:} How can user--agent experiences be formalized into computational mechanisms that sustain cross-temporal coupling of behavior and affection?
    \item \textbf{RQ2:} How does cross-temporal emotional modeling influence users’ perceptions of coherence and psychological harmony (i.e., perceived emotional resonance and alignment in human–agent interaction) during everyday interactions?
    \item \textbf{RQ3:} What design tensions characterize long-term human--agent companionship, and how do they shape user experience in real-world settings?
\end{itemize}
Through this study, we show how cross-temporal mechanisms shape daily interactions in ways that enhance perceived coherence and harmony, while also exposing the underlying tension between stability and variability in companionship. 
\paragraph{\textbf{Contributions.}} Our contributions are as follows:
\begin{itemize}
    \item \textbf{A Cross-Temporal Emotional Modeling framework for agents.} We introduce CTEM, the first computational framework that formalizes how long-term behavioral accumulation dynamically shapes agents’ affective states, and how these evolving states, in turn, condition momentary expressions and future behaviors.
    \item \textbf{A companion agent instance built upon CTEM.} We instantiate CTEM in \textit{Auri}, a companion agent, and evaluate it in a 21-day deployment. Results show that by leveraging cross-temporal accumulation to drive emotion and behavior, \textit{Auri} improves users’ perceived coherence and harmony in everyday companionship.
    \item \textbf{Design implications for future agent systems.} Drawing on findings from this deployment, we identify key tensions, such as stability versus variability and continuity versus adaptability, that shape companion-like interactions. These insights go beyond CTEM and can guide the design of affective companion agents.
\end{itemize}

\section{Background and Related Work}
\subsection{Background}

Psychological research shows that emotions unfold dynamically over time, shaped by memory, appraisal, and context rather than as isolated reactions~\cite{frijda1986emotions,mcclay2023dynamic}. Higher-order emotions such as reflection, anticipation, and regret arise from evaluating past outcomes, projecting futures, and considering counterfactuals~\cite{ocarroll2013cognitive,baumeister2007psychology}, with studies confirming such temporal evaluations as central to experience~\cite{Zeelenberg1999,Loewenstein2001,Mellers1997,Loomes1982}. Neuroscience likewise demonstrates that cognition and emotion are inseparable: they interact within shared neural systems~\cite{Ochsner2007}, integrate across prefrontal–limbic networks~\cite{Pessoa2008}, and function as dynamic value signals guiding attention, memory, and goal pursuit~\cite{Dolan2002,Pessoa2009}. Motivational theories add that needs and goals sustain behavior over time, from layered priorities~\cite{maslow1943theory} to autonomy, competence, and relatedness~\cite{ryan2000self,deci2013intrinsic} and expectancy-value models of future outcomes~\cite{wigfield2000expectancy}. Together, these perspectives highlight affect as a temporally extended process, motivating our Cross-Temporal Emotional Modeling (CTEM) framework for sustainable companionship.

\subsection{Related Work}

\paragraph{\textbf{Foundations of Human-Agent Interaction.}}  
Early research advanced engagement through socially expressive behaviors~\cite{cassell2000embodied,bickmore2005long} and through personality modeling and long-term memory architectures~\cite{park2023generative_agents,liu2024compeer}. These approaches made agents appear more natural and persistent, yet they often operationalize engagement as isolated techniques. What remains underexplored is how long-term affective accumulation shapes agents’ moment-to-moment sense of psychological harmony, and conversely, how immediate cues recursively alter enduring affective trajectories. This gap highlights the lack of models that treat affect as temporally coupled rather than session-bound.  

\paragraph{\textbf{LLM-Driven Human-Agent Interaction.}}  
Large Language Models (LLMs) now support rich social and emotional contexts~\cite{10.1145/3744746,Durante2024AgentAS}. Supportive, non-judgmental styles foster disclosure~\cite{skjuve2021my,brandtzaeg2022my,hu2025makes,yin2024ai}, while anthropomorphic cues shape rapport~\cite{qi2025assistant,pentina2023exploring,natarajan2020effects,rapp2021human}. Yet most agents still focus on short-term naturalness; memory and persona modules~\cite{park2023generative_agents,wang2025user} improve recall but rarely capture evolving emotional states. Even recent work such as~\cite{croissant2024chain_of_emotion} primarily emphasizes the generation and expression of emotions across multi-turn dialogues, while still overlooking the bidirectional dynamics. Similarly,~\cite{pataranutaporn2024future} highlights the importance of projecting future-oriented self-states in human–AI interaction, yet it does not explicitly model how an agent’s emotional traits evolve across time and recursively affect its behavioral choices. Together, these gaps underscore the need for a cross-temporal emotional modeling framework that integrates both conversational and non-conversational dynamics into the design of companionable agents. As a result, temporal coherence remains fragile, and interactions risk becoming repetitive or shallow. This motivates models that explicitly link harmony with longer-term affective dynamics.  

\paragraph{\textbf{Human-Agent Interaction via Explicit Social Behaviors.}}  
Agents frequently rely on explicit behaviors such as greetings, humor, or empathy to build rapport~\cite{cassell2000embodied,bickmore2005long,yuan2025don}. Proactive reminders and personalized small talk~\cite{skjuve2021my,brandtzaeg2022my} enhance trust, and longitudinal studies show continuity of such behaviors improves adherence in health contexts~\cite{bickmore2006health,bickmore2010maintaining}. Yet these cues are typically designed as session-bound tactics, enhancing sociability without reflecting the longer-term accumulation of users’ affective states. Compared with LLM-driven mechanisms, explicit behaviors demonstrate the value of continuity, but still lack integration with broader temporal affective dynamics.  

\paragraph{\textbf{Human-Agent Interaction via Implicit Attributes.}}  
Beyond explicit behaviors, implicit attributes such as personality and cognitive strategies shape user experience. Stable identity traits foster immersion in education~\cite{kim2021like,beege2023emotional}, services~\cite{zhang2024emotional,park2025chatbots,borau2021most,olafsson2020motivating}, and healthcare~\cite{easton2019virtual,anisha2024evaluating}. Positive dispositions enhance fluency~\cite{jun2025exploring}, and compatibility supports personalization~\cite{tu2023characterchat}. Personas further improve outcomes in sensitive domains like mental health~\cite{pataranutaporn2024future,fang2025leveraging}. At the cognitive level, memory and planning modules improve disclosure~\cite{jo2024understanding}, engagement~\cite{liu2024compeer}, and reasoning~\cite{kim2025bleacherbot,lee2025irl}. However, implicit traits often lack stability: personalities drift, memories reset, and emotional states fail to evolve across encounters. While implicit traits add depth, they, like explicit cues, fall short of sustaining affective accumulation across time.  

\paragraph{\textbf{Risks of Deep Emotional Companionship with LLMs.}}  
With digitalization reducing face-to-face contact, demand for emotional companionship is rising~\cite{gruber2022value,chaturvedi2023social}. AI chatbots mitigate loneliness and stress~\cite{li2024finding,hu2023social,mirowska2023sweet}, but deep emotional bonds risk dependency and erosion of real-world relationships~\cite{xie2023friend,adam2025supportive,fang2025ai}. \mycolor{Increasing levels of \emph{AI anthropomorphism} can amplify such risks, particularly among individuals with stronger social-connection needs~\cite{guingrich2025longitudinal}. Moreover, deep emotional bonds may emerge unintentionally even with general-purpose assistants~\cite{pataranutaporn2025my}. 
\recolor{Inspired by prior discussions on anthropomorphism and emotional attachmentt,
we intentionally adopt lightweight and non-human-oriented design choices to mitigate the risk of excessive attachment.}
This tension highlights the need for designing human–AI interaction patterns that support continuity while simultaneously incorporating safeguards to prevent users from developing excessive emotional attachment to anthropomorphized AI, which may lead to unintended negative consequences.}

\paragraph{\textbf{Summary.}}  
Taken together, prior work demonstrates the value of expressivity, memory, and personality in sustaining interaction, yet lacks formal integration of harmony with long-term affective accumulation. Our work addresses this gap by introducing \emph{Cross-Temporal Emotional Modeling (CTEM)}, a framework that explicitly couples short-term emotional harmony with long-term trajectories to enable more natural and companionable virtual agents.

\section{Agent "Auri": Design and Implementation}
\subsection{Design Objectives}  
\label{sec:method:design_objectives}
We developed \textbf{\textit{Auri}}, a foundation-model-based companion agent deployed on an instant-messaging platform, to examine how computational mechanisms can sustain cross-temporal coupling of behavior and affection in companion agents (\textbf{RQ1}).
\textit{Auri} is designed to track affect and sustain engagement via three goals:
\begin{itemize}
    \item \textbf{DG1:} Supporting open-ended, drop-in daily use with diverse contexts; users can engage at will, and idle periods should not impair later interactions.
    \item \textbf{DG2:} Using Cross-Temporal Emotion Modeling (CTEM) to link accumulated behavioral history with evolving emotional states, fostering coherent and harmonious companionship.
    \item \textbf{DG3:} Balancing stability and flexibility—adapt to user feedback while maintaining a consistent companion identity over time.
\end{itemize}
We first overview the system architecture (\cref{sec:method:systemoverview}), then detail the CTEM framework (\cref{sec:method:ctem}) in the next sections.

\subsection{System Overview}
\label{sec:method:systemoverview}
\begin{figure*}[ht]
    \centering
    \includegraphics[width=1.0\linewidth]{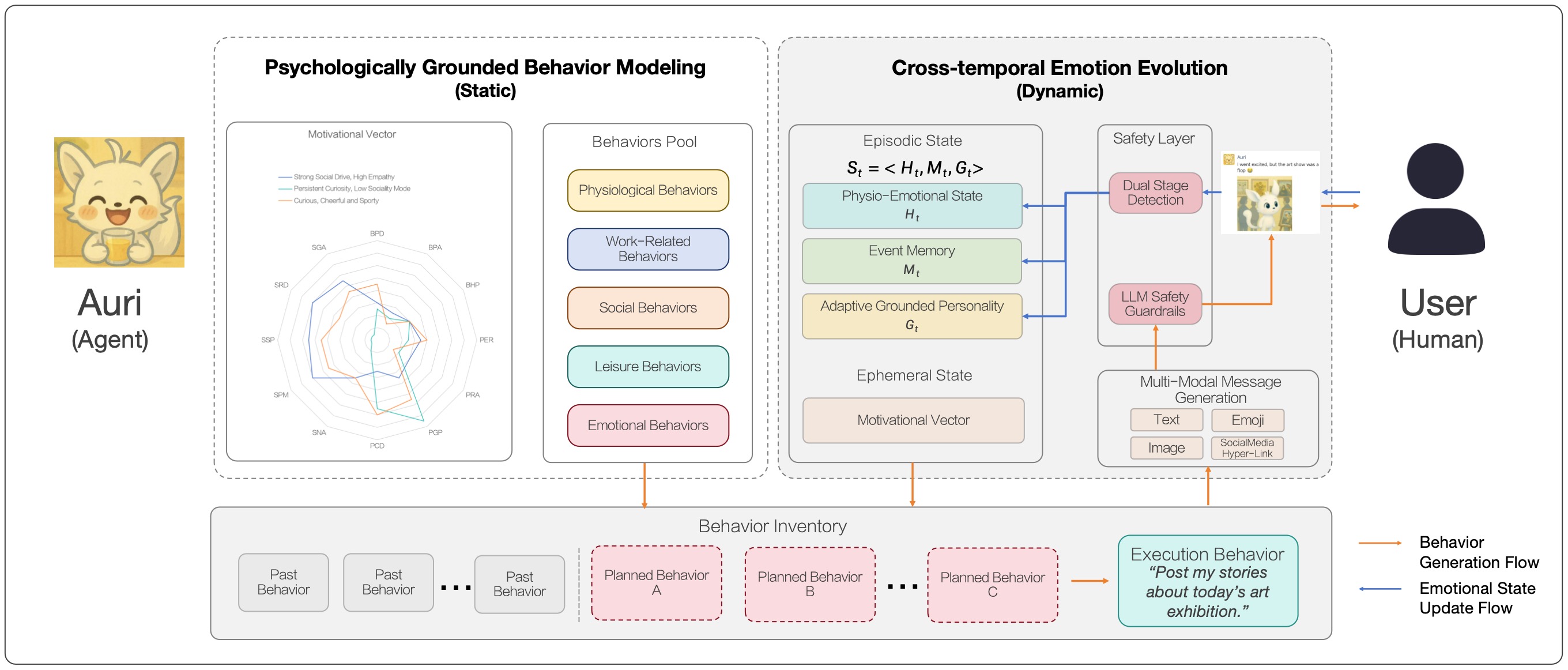}
    \vspace{-0.6cm}
    \Description{}
    \caption{System overview of Auri, showing the CTEM framework with front-end and back-end components.}
    \label{fig:SystemOverview}
\end{figure*}

\Cref{fig:SystemOverview} illustrates the overall system architecture of \textit{Auri}. 
It comprises a front-end for user interaction and a back-end driving the behavioral and emotional dynamics via the CTEM framework.


\paragraph{\textbf{Front-end.}}
\textit{Auri} is instantiated on an instant-messaging (IM) platform, utilizing its familiar, low-barrier nature to support long-term, cross-temporal affective interaction within users' daily routines.
Interacting with \textit{Auri} resembles chatting with a regular contact: it initiates greetings and contextually relevant conversations using text, emojis, and images.
Beyond direct chat, \textit{Auri} also proactively shares life updates akin to social media stories, reflecting its internal state which users can engage with via likes, comments, and reactions.

\paragraph{\textbf{Back-end.}}
\textit{Auri} is powered by the CTEM framework, which models the agent's emotional state and behavior dynamics via three core components: (1) behavior generation and integration, (2) adaptive interaction with safeguard control, and (3) emotional state updating.
By continuously updating the internal emotional state based on accumulated interactions, CTEM ensures long-term behavioral and emotion coherence while responsive to user feedback. We detail this framework in \cref{sec:method:ctem}.

\begin{table*}[ht]
\centering
\small
\renewcommand{\arraystretch}{1.0}
\begin{tabular}{cp{10cm} p{3cm} p{2cm}}
\toprule
\textbf{Variable} & \textbf{Description} & \textbf{Range / Format} & \textbf{Default} \\
\midrule
$H_t$ & Physio-emotional state: $\langle h^{phy}_t, h^{val}_t, h^{aro}_t \rangle$ capturing energy, valence, arousal & $[0,1] \times [-1,1] \times [0,1]$ & $(0.5,0.0,0.5)$ \\
\midrule
$V_t$ & Motivational drives: $\langle v^{bio}, v^{psy}, v^{soc} \rangle$ covering biological, psychological, social needs & $\mathbb{R}^{12}$, normalized & $[0.5]^{12}$ \\
\midrule
$M_t$ & Memory (self, user, context) & dynamic store & empty \\
$G_t$ & Personality modulated by $H_t$ & text prompt & baseline $G$ \\
$P$ & Behavior pool (5 categories) & structured list & predefined \\
$B_t$ & Behavioral inventory $\langle Past_t, Present_t, Future_t \rangle$ & dynamic structure & empty \\
\bottomrule
\end{tabular}
\captionsetup{width=0.9\linewidth}
\caption{CTEM variables and structures in \textit{Auri}. $H_t$ condenses three affective dimensions, and $V_t$ groups motivational drives into biological, psychological, and social categories.}
\label{tab:ctem-complete}
\end{table*}

\subsection{Cross-Temporal Emotional Modeling} 
\label{sec:method:ctem}
The \textbf{Cross-Temporal Emotional Modeling (CTEM)} framework powers \textit{Auri} by formalizing the reciprocal dynamics between long-term behavioral accumulation and evolving emotional states. 
At each timestamp $t$, CTEM generates a behavioral inventory $B_t$ from predefined categories $P$, executes one behavior (from the inventory) based on the current emotional state $S_t$, and updates both $S_t$ and $B_t$ based on the outcome and user feedback.

\paragraph{\textbf{Emotional States.}}
The emotional state acts as the core determinant of future behaviors and interaction styles. 
Inspired by motivational psychology, we formalize the state at timestamp $t$ as a tuple $S_t = \langle H_t, V_t, M_t, G_t \rangle$:
\begin{itemize}
    \item \textit{Physio-emotional state}  $H_t = \langle h^{phy}_t, h^{val}_t, h^{aro}_t \rangle$: Models physical energy(vitality), emotional valence (positivity/negativity), and arousal (calmness/excitation).
    \item \textit{Motivational vector} $V_t = \langle v^{bio}, v^{psy}, v^{soc} \rangle$: A 12-dim vector that groups biological (e.g., health), psychological (e.g., curiosity), and social (e.g., altruism) drives.
    \item \textit{Memory} $M_t$: A time-indexed store enabling chronological organization and cross-session retrieval of semantic and affective context.
    \item \textit{Adapted personality} $G_t=f(G, H_t)$: Modulates the baseline personality $G$ via state-dependent tone labels (e.g., \textit{``tired''}) injected into model prompts.
\end{itemize}
These states integrates affect, motivation, memory, and personality, ensuring the agent remains long-term consistent yet contextually adaptive. Detailed variable definitions are provided in Table~\ref{tab:ctem-complete}.

\paragraph{\textbf{Behavior Generation and Integration.}}
We introduce a \textit{Behavioral Inventory} $B_t = \langle Past_t, Present_t, Future_t \rangle$ to organize actions into executed history, immediate tasks, and future plans. 
Each entry is a tuple $\langle b, \delta_s \rangle$ consists of a behavior $b$ (e.g., go to a concert, study, rest, etc.,) and its expected state update $\delta_s$. Behaviors are not arbitarily generated; they are sampled from a pre-defined categories $P$ covering physiological, work, leisure, social, and emotional domains.
These five behavioral domains are inspired by motivational distinctions identified in affective and behavioral sciences~\cite{Krueger2009, fancourt2021, matud2024, gershuny2011time} and those exact texonomy of events is generated by synthesizing empirical daily activity patterns that belong to corrsponding domains. Each element uses a standardized format for seamless expansion (please refer to Appendix Table~\ref{tab:ctem-schema}).

The behavior selection is driven by a score function $\mathcal{S}(b)$ that evaluates each candidate behavior w.r.t current emotional state $S_t$:
\begin{flalign}
&\hspace*{1cm}
\langle v^{bio}, v^{psy}, v^{soc} \rangle 
= \phi(b) && \\[4pt]
&\hspace*{1cm}
\mathcal{S}(b) 
= w(h^{phy}_t)\,\mathrm{dot}(v^{bio}, b^{bio}) && \nonumber\\
&\hspace*{1.5cm}
\quad + \bigl(1-w(h^{phy}_t)\bigr)\,
\mathrm{dot}\bigl(
  \langle v^{psy}, v^{soc}\rangle,
  \langle b^{psy}, b^{soc}\rangle
\bigr) &&
\end{flalign}
where $\langle b^{bio}, b^{psy}, b^{soc} \rangle = \phi(b)$ is the feature embedding of behavior $b$ in the motivational space, and $w(h^{phy}_t)=1-\sigma(h^{phy}_t)$ is a modulation weight from physical energy $h^{phy}_t$. This score function captures behavior-emotion relevance while dynamically balances drives: low physical energy prioritizes biological needs, while high energy shifts focus toward psychological and social exploration. The modulation weight is bounded by the logistic function $\sigma(\cdot)$ to simulate the "diminishing returns" of emotional dynamics (i.e., reduced sensitivity near saturation).

During each behavior selection step, CTEM first plans the future steps, refilling $Future_t$ with top-scored candidates from $P$ whenever it empties. The immediate behaviors are then sampled from $Present_t$ for execution, using up-to-date scores as the target probability distribution.
This decoupling of planning and execution ensures responsiveness: if the evolving state $S_t$ makes planned behaviors unsuitable at execution time (e.g., scores drop below a threshold due to fatigue), the system triggers immediate replanning. Finally, candidates are filtered against $Past_t$ to prevent redundancy and preserve contextual consistency. (See Appendix for detailed algorithms).

\paragraph{\textbf{Adaptive Interaction.}}  
\textit{Auri} achieves adaptive interaction by integrating \textit{real-time interaction modulation} and \textit{feedback-driven adaptation} with its internal state $S_t$ and the next behavior from $Present_t$.

\paragraph{\noindent{\textit{Real-time Interaction Modulation.}}}
Selected behaviors are instantiated by a foundation model into narrative or multimodal outputs (e.g., diary entries, images) and recorded in the inventory.
By design, \textit{Auri} interacts via an instant-messaging interface, remaining accessible even during internal activity execution. Its communication style is dynamically modulated by $S_t$ (incorporating $H_t$, $M_t$, $G_t$) and real-world context (e.g., weather, time, holidays), allowing it to proactively initiate topics or reactively respond with appropriate affective tones (Appendix Algorithm ~\ref{lst:collect-feedback}, ~\ref{lst:adaptive-interaction}, and ~\ref{lst:character-prompt}).
As memory and state accumulate, the agent develops richer affective qualities such as reflection and anticipation.
To maintain coherence and mitigate LLM context window limitations, conversations are summarized and grouped by time and topic, preserving emotional consistency within interactions (Appendix Algorithm ~\ref{lst:dialog-cluster}).

To ensure trust and ethical safety, \textit{Auri} is explicitly framed as a companion rather than a romantic partner. This is supported by a dual-stage monitoring pipeline (keyword screening combined with an ensemble of LLM classifiers) that detects extreme emotions, danger signals, or over-dependence.
Upon detection, the system dynamically injects safety constraints into generation prompts, prioritizing reassurance, reminders of human agency, and professional referrals (Appendix Algorithm ~\ref{lst:safety}).
All outputs are further bounded by a safety space $\mathcal{G}_{safe}$ to ensure content safety (filtering violent, sexual, or biased material) and emotional safety (supportive rather than harmful responses)(Appendix Algorithm ~\ref{lst:basic-dialog-rule}).
Additionally, the interaction strategy adapts to user affection, ensuring that user well-being remains the primary design objective: it will priortizes active listening during distress, while engages deeper dialogue in positive moods (Appendix Algorithm ~\ref{lst:state-prompt}).

\paragraph{\noindent{\textit{Feedback-Driven Adaptation.}}} 

As interactions accumulate, \textit{Auri} rebalances $H_t$ for long-term emotional stability. Each executed behaviors triggers state updates $\delta_s$, incrementally modifying $H_t$ and $G_t$. For example, energy-demanding tasks will reduce vitality $h^{phy}_t$, while positive interactions will boost valence $h^{val}_t$. Updates are generally gradual to ensure continuity, but key restorative activities (e.g., meals, sleep) can cause larger shifts. A mandatory nightly rest phase restores energy and emotional balance to prevent system overload.
It also gradually shifts personality $G_t$ from unfamiliarity to familiarity (social relatedness $0 \to 1$), ensuring continuous evolution and deepening engagement.
Finally, \textit{Auri} also updates memory at the end of every day by incorporating new episodes via 
\[
M_{t+1} = \operatorname{UpdateMemory}(M_t, e_t).
\]

\paragraph{\textbf{Examples of Emotional-State Propagation.}}

\begin{figure*}[t]
    \centering
    \includegraphics[width=\textwidth]{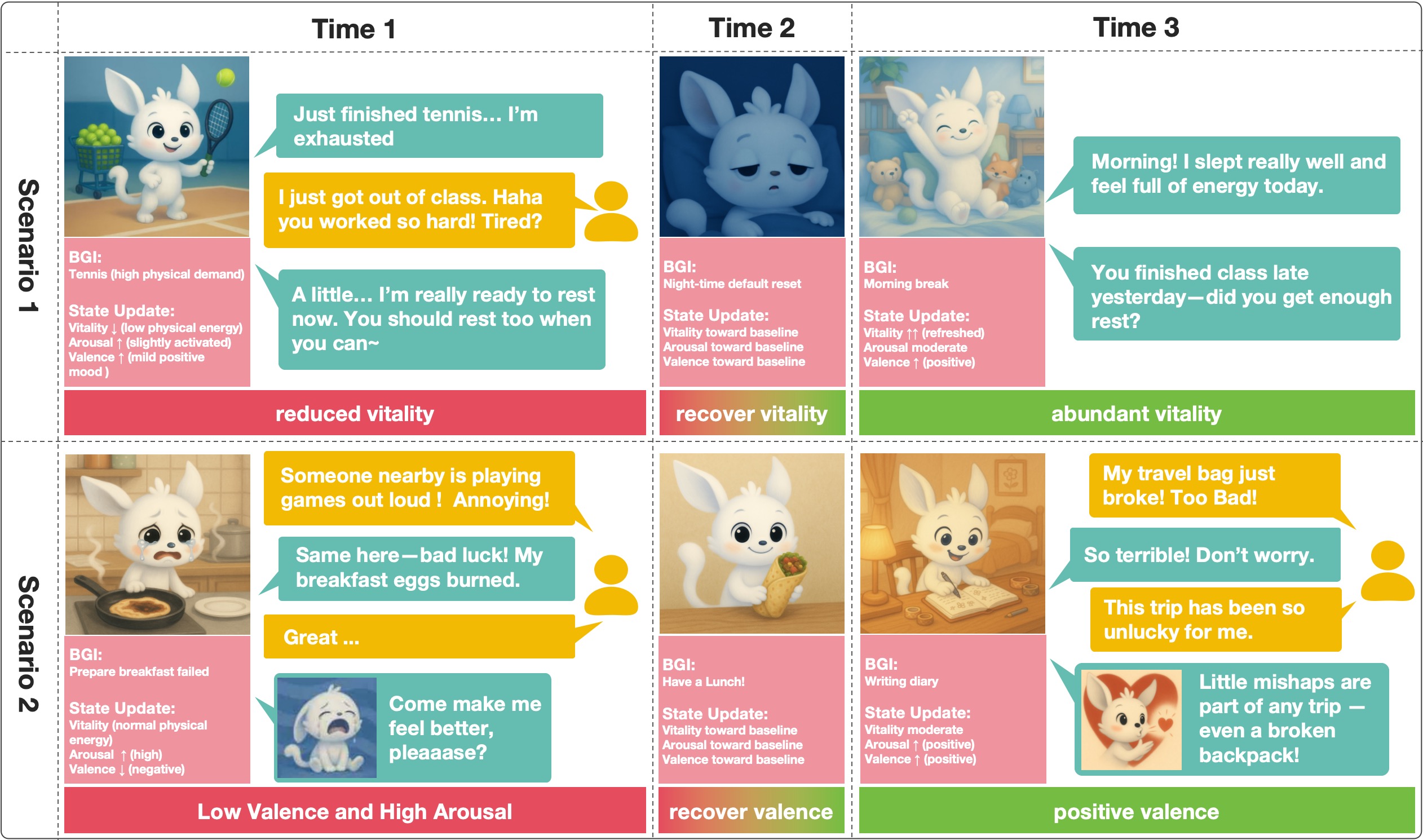}
    \captionsetup{width=0.9\textwidth}
    \Description{}
    \caption{Examples of cross-temporal interaction under CTEM across two scenarios. BGI refers to Behavior Generation and Integration, and State Update refers to Emotion State Updating.}
    \label{fig:scenario_example}
\end{figure*}

We demonstrate CTEM's cross-temporal dynamics through two concrete examples as illustrated in Figure~\ref{fig:scenario_example}.
In example 1, \textit{Auri} completes a high-effort activity which reduces its vitality (Time 1), leading to a subdued tone while planning restorative actions.
After a nightly rest (Time 2) with a restorative behavior (Time 3), \textit{Auri} regains energy, initiates next conversation with a revitalized tone while referencing the previous day's context to establish continuity.
This illustrates how behaviors drive emotional state trajectories to dynamically shape \textit{Auri's} self-presentation.
In example 2, \textit{Auri} starts with a low valence and arousal emotion, leading to self-focused, withdrawal-oriented behavior (Time 1). Then a pleasant meal shifts the state toward baseline (Time 2), reducing negative bias. At Time 3, the restored state makes \textit{Auri} enables a playful, supportive response to a user's mishap. This illustrates how \textit{Auri}'s internal dynamics modulate its communicative style, regulating reactions to user distress while maintaining affective continuity.

\section{Study Design}
To investigate whether \textit{Auri} sustains cross-temporal behavioral and emotional expressions (\textbf{RQ2}), we conducted a 21-day mixed-methods in-the-wild study, comprising a 14-day main study and a 7-day sub-study across multiple experimental groups, integrating surveys and interviews. To make the research question more tractable, we further articulated \textbf{RQ2} into three subquestions:
\begin{description}
\item[RQ2a:] \textit{How does cross-temporal modeling influence users’ perception of the Agent’s coherence?} We examine whether users feel that \textit{Auri} maintains a clear and consistent internal state across time, including stable personality cues, behaviors, and emotional expressions.
\item[RQ2b:] \textit{How does cross-temporal modeling influence users’ own emotional and experiential responses?} We investigate whether users feel affected by \textit{Auri}'s behavior and expressions—such as feeling more aligned, supported, or emotionally influenced during the interaction.
\item[RQ2c:] \textit{What is the overall effect of CTEM on users’ experience and perception?} By comparing systems with and without CTEM, we evaluate the overall contribution of CTEM to users’ perception of coherence, harmony, and interaction quality in daily use.
\end{description}

\subsection{Participants}
We recruited 96 participants (38 male, 55 female, 3 non-binary; aged 18--26) from multiple universities, representing \textit{Auri}'s core target demographic. All were native speakers and regular instant messaging users, ensuring platform familiarity. Enrollment was limited to ensure sustained daily commitment, prioritizing engagement quality. Participants came from diverse academic backgrounds without specific disciplinary requirements. All participants received appropriate financial compensation upon completion.

\subsection{Experimental Platform}
We utilized a widely used mobile instant messaging app as the interaction platform. This choice leveraged users' existing familiarity with the app's interface and features, reducing the learning curve associated with adopting a new system. The app's comprehensive functionality supported rich multimodal interactions, including text, voice, images, and emojis, aligning well with \textit{Auri}'s design for naturalistic communication.

\subsection{Baseline}
As a baseline, we implemented an agent without CTEM but retaining standard chatbot capabilities, including fluent conversation, cross-session factual memory, and dialog safety. While its factual memory supported content-level continuity by recalling past information, the baseline strictly excluded temporal affective or behavioral modeling, i.e., this baseline neither track emotional states nor have behavioral evolution. This design ensured realistic long-term usability while isolating the specific affective and behavioral dynamics introduced by CTEM for analysis purposes.

\subsection{Experimental Design and Conditions}
We conducted two studies to evaluate user experience, ensuring all participants eventually interacted with the full CTEM system.

The \textbf{Main Study} ($N=54$) utilized a two-week, single-blind ablation design to isolate contributions of CTEM's core components.
In the first week (Phase 1), participants were randomly assigned to one of four groups—``Full'', ``NoMultimodal→Full'', ``NoEmo→Full'', or ``NoBehav→Full''. Each group (except the ``Full'' group) in phase 1 excluded a specific functionality (Adaptive Interaction, Emotion State Updating, or Behavior Generation \& Integration) to probe its role of emotional-expression and behavioral cues in establishing \textit{Auri}’s coherent traits (RQ2a) and emotional support to users (RQ2b). In the second week (Phase 2) the full system is enabled for all groups (Table~\ref{tab:exp-design-summary}).

The \textbf{Sub-study} ($N=42$) assessed the overall impact of CTEM through a order-controlled comparison to the baseline system. The experiment is counterbalanced; to account for differences in system complexity, participants used the baseline for 2 days and the full CTEM system for 5 days. This design minimized contrast effects~\cite{HovlandSherif1952} while capturing comparative experiential data.
All participants were randomly assigned to ``Baseline→Full'' or ``Full→Baseline'' sequences.

\begin{table*}[t]
\centering
\small
\renewcommand{\arraystretch}{1.05}
\setlength{\tabcolsep}{3pt}

\begin{tabular}{@{} l c p{3.6cm} p{3.6cm} p{5.4cm} @{}}
\toprule
\multicolumn{5}{c}{\textbf{Main Study / Ablation Study}} \\
\midrule
\textbf{Group} & \textbf{N} & \textbf{Phase 1 (7 Days)} & \textbf{Phase 2 (7 Days)} & \textbf{Description} \\
\midrule
Full & 12 & Full CTEM & Full CTEM & --- \\
NoMultimodal$\rightarrow$Full & 12 & CTEM-w/o-Multimodal & Full CTEM & Phase 1 without multimodal interactives \\
NoEmo$\rightarrow$Full & 15 & CTEM-w/o-Emotion & Full CTEM & Phase 1 without emotional expression \\
NoBehav$\rightarrow$Full & 15 & CTEM-w/o-Behavior & Full CTEM & Phase 1 without behavioral sharing \\
\midrule
\multicolumn{5}{c}{\textbf{Sub Study}} \\
\midrule
\textbf{Group} & \textbf{N} & \textbf{Phase 1} & \textbf{Phase 2} & \textbf{Description} \\
\midrule
Baseline$\rightarrow$Full & 21 & Baseline (2 Days) & Full CTEM (5 Days) & 2 days baseline followed by 5 days CTEM \\
Full$\rightarrow$Baseline & 21 & Full CTEM (5 Days) & Baseline (2 Days) & 5 days CTEM followed by 2 days baseline \\
\bottomrule
\end{tabular}

\captionsetup{width=0.95\textwidth}
\caption{Summary of experimental conditions across the main and sub studies. Groups differ in the Phase~1 system variant in the main study and in the exposure order in the sub-study.}
\label{tab:exp-design-summary}
\end{table*}

\subsection{Task and Procedure}
Our study was approved by the Institutional Review Board (IRB) of the university, in accordance with established ethical guidelines for human-computer interaction research. All participants are provided informed consent prior to participation.

\paragraph{\textbf{Onboarding.}} 
Participants completed a demographic pre-survey before accessing our system. During onboarding, they were reminded of their right to withdraw at any time and advised that \textit{Auri} is not a licensed mental health professional, with instructions to seek qualified support if distressed. Our analyses were conducted on anonymized data, with summaries generated by language models rather than human inspection. Researchers never accessed real names and no identifiers appeared in publications to ensure privacy.

\paragraph{\textbf{Experimental Procedure (Main and Sub Study).}} 
Participants engaged in long-term, in-the-wild interactions with \textit{Auri}. In the main study, participants completed brief surveys at the end of each phase to assess how their user experience changes across phases. In the Sub-study, participants completed a single comparative survey after using both system versions. A random subset of participants from both studies was selected for semi-structured interviews upon completion.



\paragraph{\textbf{Data Collection and Post-Task Interviews.}} 
To address \textbf{RQ2}, we collected data via surveys, open-ended responses, and post-phase semi-structured interviews designed to capture situated interaction experiences.

\begin{enumerate}
\item{\textbf{Survey Design.}}
The survey comprised 9 items across three dimensions:
\begin{itemize}
\item \textbf{RQ2a (Perceived Coherence):} Q1 (multiple-choice) identified observed coherent traits;  Q2 (1--5 Likert) rated their importance; Q3 (open-ended) how participants configured \textit{Auri}’s personality. 
\item \textbf{RQ2b (Perceived Harmony):} Q4 (1--4 Likert) rated response appropriateness of \textit{Auri}; Q5 (1--5 Likert) assessed emotional change after interaction; Q6 (multiple-choice) identified frequent usage contexts with \textit{Auri}.
\item \textbf{RQ2c (Overall Experience):} Q7 (1--5 rating) compared experiences across phases;. Q8--Q9 evaluated the contribution of core CTEM modules (BGI, AdI, ESU) to coherence and harmony.
\end{itemize}

Participants were administered to answer Q1-Q6 twice during the main study (during Phase 1 and Phase 2), whereas Q7-Q9 were administered once after participants had experienced both versions. Perceived coherence was measured using self-report items adapted from Antonovsky’s Sense of Coherence scale ~\cite{antonovsky1987unraveling} and prior work on continuity in interactive systems ~\cite{forlizzi2004understanding}.
Full survey questions are listed in Table~\ref{tab:survey-items}, with details in Appendix Table~\ref{tab:appendix-question-sum}.




\begin{table*}[th]
\centering
\small
\renewcommand{\arraystretch}{1.0} 
\begin{tabular}{p{2.5cm} p{1cm} p{13cm}}
\toprule
\textbf{Question} & \textbf{ID} & \textbf{Survey Item} \\
\midrule
\multirow{3}{*}{RQ2a: Coherence} 
& Q1 & In your long-term interactions with \textit{Auri}, which coherent traits of \textit{Auri} did you perceive? \\
& Q2 & To what extent did these traits make you feel that \textit{Auri} had a coherent and unified existence? \\
& Q3 & Did you feel able to adjust \textit{Auri}’s personality or behavior to better align with your preferences? \\
\midrule
\multirow{3}{*}{RQ2b: Harmony} 
& Q4 & Did \textit{Auri} provide appropriate help or responses based on your situational context? \\
& Q5 & How did your emotional state change after interacting with \textit{Auri}? \\
& Q6 & In what contexts did you most often use \textit{Auri} and expect responses? \\
\midrule
\multirow{3}{*}{RQ2c:  Overall Effect}
& Q7 & How did your usage and emotional experiences differ across \textit{Auri}’s versions? \\
& Q8 & Evaluate the contribution of each functional module to your perception of coherence. \\
& Q9 & Evaluate the contribution of each functional module to your perception of harmony. \\
\bottomrule
\end{tabular}
\caption{Survey items grouped by research question.}
\label{tab:survey-items}
\end{table*}

\item{\textbf{Qualitative and Thematic Analysis.}}
We analyzed all open-ended responses and 18 interview transcripts using inductive thematic analysis to summarize them into a set of themes. Two researchers independently coded the data, identified recurring themes, and resolved disagreements through discussion with iterative refinement of categories.

\end{enumerate}

\subsection{Analysis Approach}
We adopted mixed methods combining quantitative analyses with qualitative feedback. Effect sizes and confidence intervals are reported where applicable. Unless specifically noted, the significance level was set at $p=.05$.

For \textbf{RQ2a}, we modeled trait recognition via binomial GEE (Holm adjustment) and coherence ratings via three-way ANOVA with Bonferroni-corrected post-hoc tests; personality adjustability was analyzed through distribution patterns and qualitative accounts..

For \textbf{RQ2b}, we used cumulative link mixed models (CLMMs) (robustness checked with ordinal GEE) for harmony and emotional change, Chi-square tests for comparing usage contexts between group, and McNemar’s tests for within-participant phase changes.

For \textbf{RQ2c}, retrospective usage improvement over baseline were assessed with one-sample t-tests (with Wilcoxon tests as robustness checks), and comparisons of emotional scores between baseline and Full phases were conducted using Wilcoxon signed-rank tests. To examine contributions of CTEM mechanisms, we applied Wilcoxon and Friedman tests for overall ratings, and CLMM to model condition effects, with GEE as a robustness check.

\section{Results}
\subsection{RQ2a: CTEM Strengthens \textit{Auri}'s Coherence}

This section reports main study results (Q1–Q3) on perceived coherent traits, their importance, and personality adjustability.

\paragraph{\textbf{Perception of Auri's Coherent Traits.}}
In Q1, participants identified which coherent traits they perceived in \textit{Auri} during each phase. Binary selections were analyzed using a binomial GEE. Table~\ref{tab:positive_trait_probs_ci} shows the estimated overall probabilities for positive traits, with details in Appendix Table~\ref{tab:appendix_coherent_traits}.

\begin{table}[ht]
\centering
\small
\setlength{\tabcolsep}{6pt} 
\begin{tabular}{p{3cm} cc cc}
\toprule
\multicolumn{1}{c}{\textbf{Group}}
& \multicolumn{2}{c}{\textbf{Phase 1}} 
& \multicolumn{2}{c}{\textbf{Phase 2}} \\
\cmidrule(lr){2-3} \cmidrule(lr){4-5}
& $p$ & 95\% CI & $p$ & 95\% CI \\
\midrule
NoMultimodal$\rightarrow$Full & .917 & {[}.587,.999{]} & .999 & {[}.999,.999{]} \\
NoBehav$\rightarrow$Full      & .867 & {[}.595,.999{]} & .915 & {[}.525,.999{]} \\
NoEmo$\rightarrow$Full        & .800 & {[}.530,.961{]} & .869 & {[}.439,.999{]} \\
\bottomrule
\end{tabular}
\captionsetup{width=0.95\linewidth}
\caption{Marginal probabilities and 95\% confidence intervals of perceived positive coherent traits across groups and phases.}
\label{tab:positive_trait_probs_ci}
\end{table}

Across all groups, participants endorsed positive coherent traits more frequentl than the negative option.
When ``Behavioral'' or ``Emotional'' modules were removed in Phase 1, the likelihood of perceiving coherence declined substantially, highlighting their foundational role.
In contrast, removing multimodal interaction itself had a smaller impact; high coherence ratings persisted without it. However, restoring multimodal cues in Phase 2 amplified existing coherence impressions generated with ``Behavioral'' and ``Emotional'' modules, specifically boosting user perceived appearance-related coherence (detail in Appendix).

\begin{figure}[ht]
  \centering
\includegraphics[width=\linewidth]{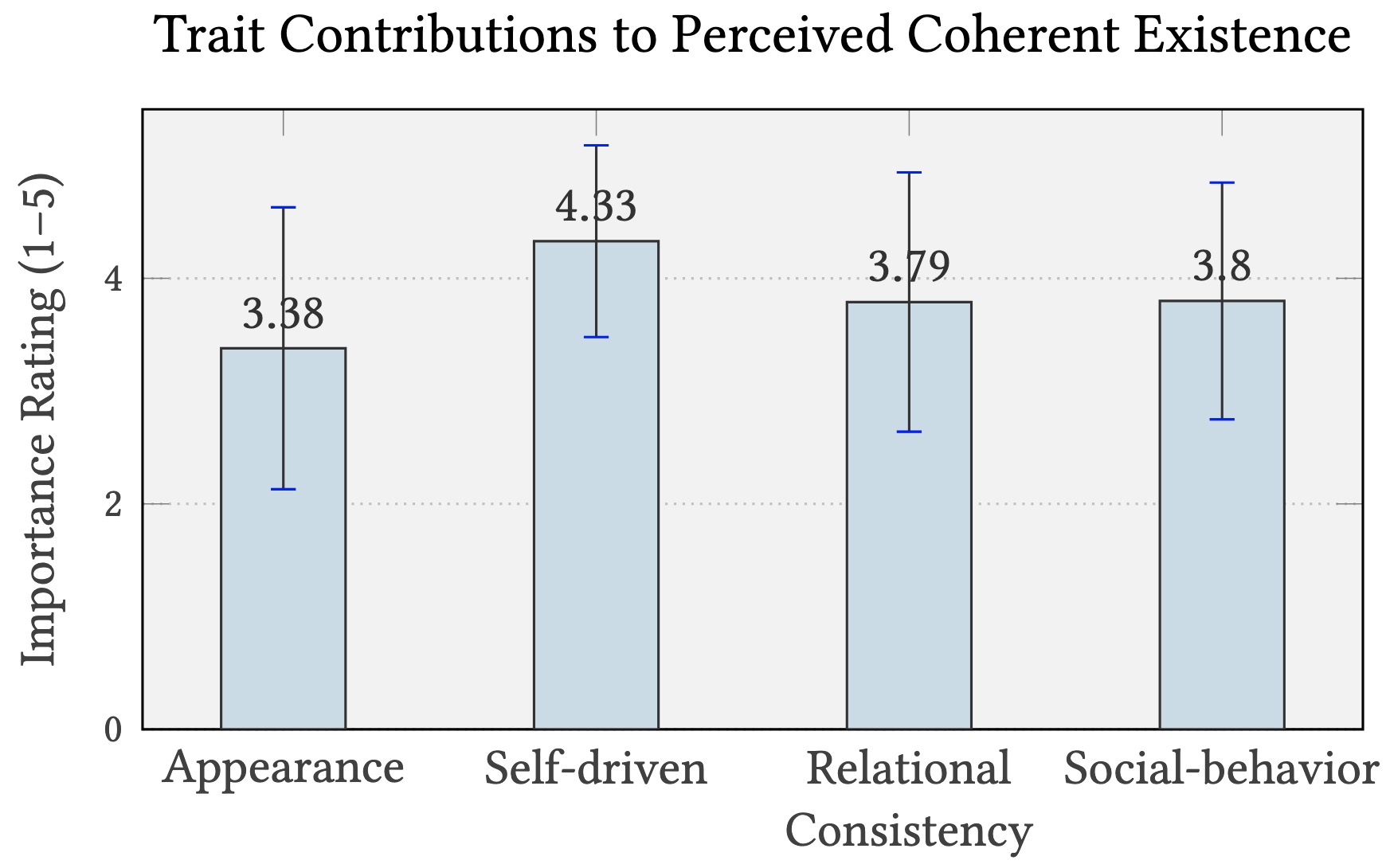}
  \vspace{-0.25cm}
  \captionsetup{width=0.95\linewidth}
  \Description{}
  \caption{Mean ratings of trait contributions to perceived coherent existence (1 = not important, 5 = very important). Self-driven refers to Self-driven\&personality-like.}
\label{fig:trait_continuity_bar}
\end{figure}

\paragraph{\textbf{Perceived Coherence of Existence.}}
In Q2, participants rated the importance of each trait for \textit{Auri}'s coherent existence (1–5 scale, 1 = not important, 5 = very important).
A three-way ANOVA revealed a significant main effect of trait ($F=15.921$, $p<.01$, $\eta^{2}=0.107$). Bonferroni-corrected Post-hoc tests indicated that ``Self-driven \& personality-like'' was the strongest contributor ($p<.01$), followed by ``Relational consistency'' and ``Social behavior'' in the mid-range, with ``Appearance'' rated lowest. Descriptive patterns were consistent with these inferential results(Fig~\ref{fig:trait_continuity_bar}).

Time and Trait×Time effects were non-significant ($p>.05$). While the ``Full'' group showed modest descriptive increases from Phase 1 (+0.25) to Phase 2 (+0.58), the relative importance ranking of traits remained stable across phases and conditions.

\paragraph{\textbf{Adjust Ability and Coherence.}} 
In Q3, 81.5\% of participants expected \textit{Auri} to have personality adjustability. Overall, participants explored an average of 2.67 personas.  Persona labels were mapped onto a two-dimensional space of warmth (high vs.\ low) and interactivity (high vs.\ low), and all four quadrants were represented. 
We observe that participants actively adjusted interaction styles to match their diverse preferences. As one participant noted: ``A companion AI is meant for companionship, so it should better fit the user’s needs.'' 

\subsection{RQ2b: Stability and Emotional Improvement in Perceived Harmony}
This section reports results from Q4–Q6 regarding perceived contextual appropriateness, emotional change, and usage patterns.

\paragraph{\textbf{Perceived Appropriateness and Emotional Change.}} 
We analyzed Q4 (contextual appropriateness) and Q5 (emotional change) using cumulative link mixed models (CLMMs; cumulative logit with random intercepts for participants), verified with ordinal GEEs.
For Q4, appropriateness ratings (1 = no appropriate support; 4 = highly appropriate support) were generally stable among phases, groups, or interactions (all $p>.26$). Exploratory CLMM-based within-group comparisons indicated a tentative increase in the NoEmo→Full'' group from Phase 1 to Phase 2 ($p=.040$). While ratings across all groups generally remained in the mid-to-upper range (Table~\ref{tab:Q4Q5_within}), the NoEmo→Full'' group exhibited a notable shift: the proportion of above-median scores rose from 20.0\% in Phase 1 to 66.7\% in Phase 2.

For Q5, participants rated the emotional change on a 1–5 scale (1= much worse; 5 =much better; 3 = no change). The CLMM showed a significant main effect of Phase ($p = .005$, $OR = 13.99$), although the confidence interval for this effect was relatively wide. No reliable group-level differences emerged. CLMM-based within-group comparisons showed that both the ``Full'' group ($p = .005$) and ``NoEmo→Full'' group ($p < .001$) exhibited clear emotional improvement from Phase 1 to Phase 2, with the ``NoEmo→Full'' group showing the largest gain. All groups had mean emotional-change ratings above 3 (Table~\ref{tab:Q4Q5_within}). 

In summary, contextual appropriateness remained stable across conditions, while emotional experience showed a generally positive change, with the largest improvement in the “NoEmo→Full” group.

\begin{table*}[t]
\centering
\small
\renewcommand{\arraystretch}{1.05}
\setlength{\tabcolsep}{6pt}

\begin{tabular}{l >{\centering\arraybackslash}p{3.0cm}
>{\centering\arraybackslash}p{2.5cm}
>{\centering\arraybackslash}p{2.5cm}
>{\centering\arraybackslash}p{3.0cm}
>{\centering\arraybackslash}p{2.5cm}}
\toprule
\textbf{Group} 
& \textbf{M $\pm$ SD (Q4, P1 $\rightarrow$ P2)} 
& \textbf{Prop. $>$2 (\%)} 
& \textbf{CLMM $p$ (Q4)} 
& \textbf{M $\pm$ SD (Q5, P1 $\rightarrow$ P2)} 
& \textbf{CLMM $p$ (Q5)} \\
\midrule

Full         
& 2.50$\pm$0.80$\rightarrow$2.67$\pm$0.49 
& 50.0$\rightarrow$66.7 
& .512      
& 3.17$\pm$0.58$\rightarrow$3.83$\pm$0.39 
& \textbf{.005} \\

NoEmo$\rightarrow$Full        
& 2.27$\pm$0.80$\rightarrow$2.67$\pm$0.72 
& 20.0$\rightarrow$66.7 
& \textbf{.040} 
& 3.20$\pm$0.56$\rightarrow$3.93$\pm$0.59 
& \textbf{$<$.001} \\

NoBehav$\rightarrow$Full      
& 2.20$\pm$0.68$\rightarrow$2.53$\pm$0.64 
& 33.3$\rightarrow$46.7 
& .188      
& 3.60$\pm$0.83$\rightarrow$3.53$\pm$0.74 
& .747 \\
\bottomrule
\end{tabular}
\captionsetup{width=0.95\linewidth}
\caption{Descriptive statistics and within-group CLMM results for Q4 (contextual appropriateness) and Q5 (emotional change) across phases. For Q4, the ``NoEmo$\rightarrow$Full'' group exhibited an increase in Phase~2. For Q5, emotional-change ratings improved in the ``Full'' and ``NoEmo$\rightarrow$Full'' groups, suggesting that restoring the emotional module may have contributed to perceived appropriateness and emotional experience.}
\label{tab:Q4Q5_within}
\end{table*}

\paragraph{\textbf{Usage Contexts.}}
In Q6, participants selected their most frequent scenarios from six common interaction contexts \mycolor{
(Appendix Table~\ref{tab:usage_contexts_definition})}.
We used chi-square tests to assess between-group differences and McNemar’s paired tests to evaluate phase-to-phase changes within participants.

Chi-square tests showed no between-group differences in either phase (Phase 1: all $p\ge.25$; Phase 2: all $p\ge.49$), indicating broadly comparable usage patterns across the four groups.
In Phase 1, ``Fragmented Time'' (34.3\%) and ``Emotional Exchange'' (30.5\%) were the most common contexts, while deeper engagement contexts such as ``Emotional Bonding'' (10.5\%) and ``Support \& Care'' (8.6\%) were relatively infrequent.
In Phase 2, deeper engagement contexts remained less frequent than lightweight social contexts, which continued to dominate participants’ interactions. Notably, users’ specific usage preferences within lightweight contexts shifted over time.

\begin{figure}[ht]
\centering
\includegraphics[width=\linewidth]{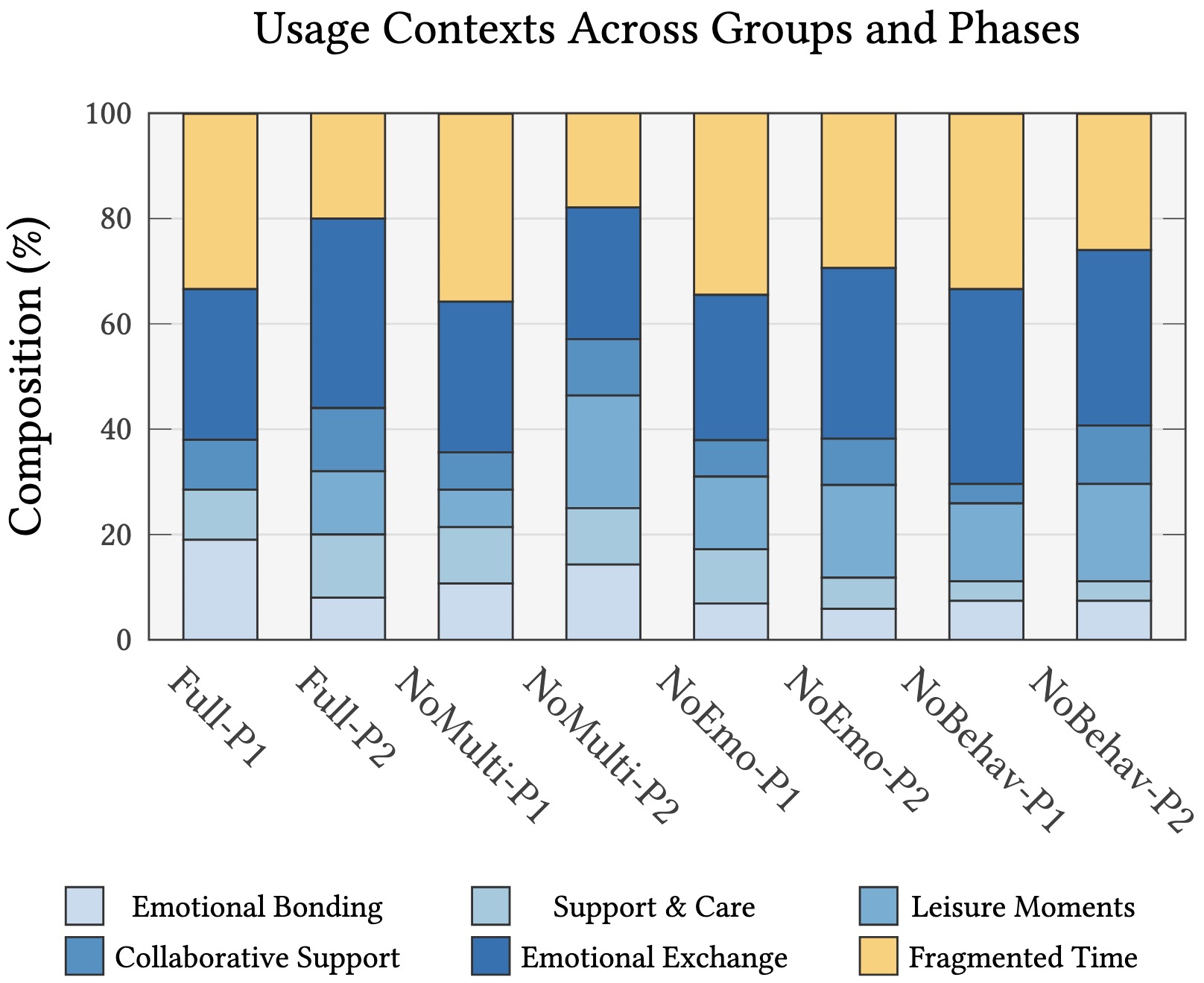}
\vspace{-0.25cm}
\captionsetup{width=0.95\linewidth}
\Description{}
\caption{Distribution of usage contexts across groups in Phase 1 and Phase 2. Each bar shows the percentage of participants selecting each context within a group-phase combination. 
Lightweight social contexts contexts, such as ``Emotional Exchange'' and ``Leisure Moments'' were more frequently selected, whereas deeper emotional engagement contexts were selected less frequently.}
\label{fig:usage_contexts}
\end{figure}

We conducted McNemar's paired tests to assess the significance of changes in lightweight social contexts between Phase 1 and Phase 2. ``Leisure Moments'' increased significantly ($\chi^2(1)=5.56$, $p=.018$), and ``Fragmented Time'' decreased significantly ($\chi^2(1)=3.86$, $p=.050$). This pattern was consistent across all four groups (Fig.~\ref{fig:usage_contexts}).

Overall, participants moved from fragmented use toward more intentional, context-specific and routine-integrated use, while emotionally demanding contexts remained consistently low.  

\subsection{RQ2c: Perceived Coherence and Harmony under Different Configurations}
This section presents sub-study results (Q7–Q9) on usage and emotional experiences, as well as the contributions of CTEM mechanisms to coherence and harmony.

\paragraph{\textbf{Usage and Emotional Experience.}}
In Q7, participants rated usage experience improvement (1=worse, 2=no change, 5=significantly better). 
We analyzed these ratings using one-sample $t$-tests against the ``no change'' benchmark, with Wilcoxon tests as a robustness check.
One-sample $t$-tests confirmed significant improvement over the ``no change'' benchmark for all groups (all $p < .001$, Table~\ref{tab:Q7}), with the largest gain in the ``Baseline$\rightarrow$Full'' group. When against a stricter benchmark of 3 (``slight improvement''), the overall improvement ($p = .041$) and the ``Baseline$\rightarrow$Full'' group improvement ($p = .008$) remained significant, while the ``Full$\rightarrow$Baseline'' group did not ($p = .680$).

For emotional experience (1=much worse, 3=no change, 5=much better), ratings improved significantly from ``Baseline'' ($M = 3.24$) to ``Full'' ($M = 3.67$) ($W=32, p = .004, r = .53$). Within-group analyses confirmed this positive shift in both order conditions (all $p < .05$, Table~\ref{tab:Q7}).

\begin{table}[ht]
\centering
\small
\renewcommand{\arraystretch}{1.05}
\setlength{\tabcolsep}{4pt}

\begin{tabular}{p{2cm} c c c c}
\toprule
\multicolumn{5}{c}{\textbf{Usage Experience (vs.\ 2 = no change)}} \\
\midrule
\textbf{Group}
& \textbf{M$\pm$SD}
& \textbf{$t$}
& \textbf{$p$}
& \textbf{$d$ / Wilcoxon $p$} \\
\midrule
Baseline$\rightarrow$Full   
& 3.52$\pm$0.81 & 8.58 & $<.001$ & 1.87 / $<.001$ \\
Full$\rightarrow$Baseline   
& 3.10$\pm$1.04 & 4.81 & $<.001$ & 1.05 / .001 \\
Total    
& 3.31$\pm$0.95 & 8.94 & $<.001$ & 1.38 / $<.001$ \\

\midrule
\multicolumn{5}{c}{\textbf{Emotional Experience (vs.\ 3 = no change)}} \\
\midrule
\textbf{Group}
& \textbf{M$\pm$SD (B)}
& \textbf{M$\pm$SD (F)}
& $\Delta$
& \textbf{Wilcoxon $p$} \\
\midrule
Baseline$\rightarrow$Full   
& 3.43$\pm$0.51 & 3.71$\pm$0.46 & +0.29 & .034 \\
Full$\rightarrow$Baseline   
& 3.05$\pm$0.92 & 3.62$\pm$0.92 & +0.57 & .030 \\
Total    
& 3.24$\pm$0.67 & 3.67$\pm$0.72 & +0.43 & .004 \\
\bottomrule
\end{tabular}
\captionsetup{width=0.98\columnwidth}
\caption{Usage experience and emotional change across system versions. Usage statistics are one-sample tests (vs.\ 2). Emotional statistics are paired Wilcoxon tests (Baseline vs.\ Full).}
\label{tab:Q7}
\end{table}

\paragraph{\textbf{Contributions of CTEM Mechanisms to Coherence and Harmony.}}
In Q8 and Q9, We analyzed contribution ratings (1–5, 1 = not important, 5 = extremely important, 3 = neutral) for the three CTEM mechanisms (BGI, AdI, ESU).
For coherence (Q8), all mechanisms were rated significantly above neutral ($p<.001$) as shown in Table~\ref{tab:ctem-mechanisms}. A Friedman test found significant differences ($p=.012$), with CLMM confirming that AdI and ESU received higher ratings than BGI ($p_{\text{AdI-BGI}} = .009$), as shown in Appendix Table~\ref{tab:clmm-q8-q9-appendix}.

For harmony (Q9), AdI and ESU exceeded the neutral benchmark, whereas BGI did not. While the overall Friedman test was non-significant ($p=.135$) in (Table~\ref{tab:ctem-mechanisms}), CLMM analyses similarly indicated that AdI and ESU tended to be rated higher than BGI ($p < .03$) in Appendix Table~\ref{tab:clmm-q8-q9-appendix}.

In sum, both coherence and harmony ratings suggested that AdI and ESU tended to contribute more to perceived quality than BGI.

\begin{table}[ht]
\centering
\small
\renewcommand{\arraystretch}{1.1}
\setlength{\tabcolsep}{4pt}
\begin{tabular}{
@{}
>{\centering\arraybackslash}p{0.6cm}
>{\centering\arraybackslash}p{0.8cm}
>{\centering\arraybackslash}p{1.2cm}
>{\centering\arraybackslash}p{1.0cm}
>{\centering\arraybackslash}p{1.5cm}
>{\centering\arraybackslash}p{1.5cm}
@{}
}
\toprule
\textbf{Type} & \textbf{Dim.} & \textbf{M$\pm$SD} & \textbf{Median} & \textbf{Wilcoxon $p$} & \textbf{Friedman $p$} \\
\midrule
BGI & Coh.   & 3.71 $\pm$ 0.92 & 4.0 & $<.001$ & \multirow{3}{*}{.012} \\
AdI                & Coh.   & 4.14 $\pm$  0.90 & 4.0 & $<.001$ & \\
ESU          & Coh.   & 3.90 $\pm$  0.88 & 4.0 & $<.001$ & \\
\addlinespace[2pt]
BGI & Har. & 3.33 $\pm$  1.18 & 3.5 & .060   & \multirow{3}{*}{.135} \\
AdI                & Har. & 3.45 $\pm$  0.94 & 3.5 & .003   & \\
ESU          & Har. & 3.64 $\pm$  0.85 & 4.0 & $<.001$   & \\
\bottomrule
\end{tabular}
\captionsetup{width = 0.95\linewidth}
\caption{Perceived contributions of CTEM mechanisms to coherence (Q8) and harmony (Q9), based on sub-study participants ($N=42$). AdI and ESU consistently received higher contribution ratings than BGI, indicating a stable AdI/ESU > BGI trend.}
\label{tab:ctem-mechanisms}
\end{table}

\subsection{Qualitative Findings}
Our inductive thematic analysis based on interviews from 18 participants identified five themes regarding long-term coherence and harmony (Table~\ref{tab:qual-themes}). 
Inter-coder agreement reached substantial reliability (Cohen's $\kappa=0.82$), and no new themes emerged after coding approximately 85\% of the qualitative corpus.
A summary of these themes is provided in Table~\ref{tab:qual-themes}, additional illustrative quotes are listed in the Appendix Table~\ref{tab:appendix_qualitative}.
Our qualitative analysis further explained the two major patterns identified in the quantitative analysis, confirming the value of our CTEM.

\begin{table*}[ht]
\centering
\small
\renewcommand{\arraystretch}{1.2}
\setlength{\tabcolsep}{4pt}
\centering
\begin{tabular}{p{4cm}p{6cm}p{5cm}p{1cm}}
\toprule
\textbf{Theme} & \textbf{Definition} & \textbf{Representative Quote} & \textbf{\#P} \\
\midrule
Memory Continuity &
AI recalls past facts, preferences, and conversations &
``It remembered what I said days ago.'' &
17 \\

Stable Persona \& Tone &
Consistent linguistic, personality, and emotional style &
``It keeps a lively, consistent tone.'' &
14 \\

Life Narrative Continuity &
Maintains ongoing daily storyline (events, moods) &
``It followed up on yesterday’s topic.'' &
10 \\

Breaks Continuity (Control Group) &
Forgetfulness or abrupt topical/emotional shifts &
``The shift from sad to happy was jarring.'' &
16 \\

Emotional Harmony &
Emotionally attuned, supportive responses & ``When I felt anxious, it supported me.'' &
15 \\
\bottomrule
\end{tabular}
\vspace{0.2cm}
\caption{Key themes from the inductive thematic analysis.}
\label{tab:qual-themes}
\end{table*}

\textbf{Strengthened long-term coherence.}
Long-term coherence was characterized by three positive themes and one disruption theme. ``Memory Continuity'' was central to identity perception: when \textit{Auri} recalled past conversations and preferences, participants experienced it as ``the same being'' across sessions. ``Stable Persona \& Tone'' reinforced this sense through consistent linguistic and emotional style. 
``Life Narrative Continuity'', reflected in ongoing storyline and follow-ups, fostered a sense of shared trajectory rather than isolated moments.
Conversely, ``Breaks in Continuity'' (Control group) posed the primary threat to coherence in the absence of CTEM. Sudden forgetfulness or abrupt topical and emotional shifts disrupted the sense of engaging with a stable entity, as reported by 89\% of participants.

\textbf{Enhanced harmony.}
The theme ``Emotional Harmony'' highlighted interaction quality, with emotionally attuned and contextually appropriate responses fostering feelings of being understood and supported.

These observations point to two higher-level mechanisms. 
First, affective carry-over enabled \textit{Auri} to retain emotional and behavioral context through consistent memory and persona cues, effectively reducing the perception of a ``reset'' between interactions. 
Second, anticipatory alignment ensured that \textit{Auri}’s responses were timely and emotionally attuned to users’ needs, resulting in interactions that were both contextually appropriate and supportive.
Taken together, CTEM not only reduced the common ``reset'' effect but also allowed \textit{Auri} to present more complex emotional features and behaviors.
Such expressions made \textit{Auri} appear more consistent and more sensitive to context.
These mechanisms explain the most frequently endorsed coherent traits in Q1 and Q2 as well as the Phase-wise improvement in Q5.
Overall, by bridging interaction gaps with complex emotional behaviors, CTEM enhances coherence (\textbf{RQ2a}) and harmony (\textbf{RQ2b}), and fulfill \textbf{DG1} by ensuring that pauses in use do not disrupt subsequent experiences.

Finally, we constructed a \textit{joint display} that aligns key quantitative effects with inductively derived qualitative themes (Table~\ref{tab:joint-display}). This triangulation clarifies why CTEM yields the observed gains and when such gains are more likely to surface, thereby fulfilling \textbf{DG2} by linking long-term affective accumulation with evolving emotional states to ensure coherent and harmonious companionship.

\begin{table*}[ht]
\centering
\small
\renewcommand{\arraystretch}{1.2}
\begin{tabular}{p{4.5cm} p{6.5cm} p{5cm}}
\toprule
\textbf{Quantitative Finding} & \textbf{Qualitative Theme} & \textbf{Design/Interpretive Takeaway} \\
\midrule
Higher selection of positive trait ``Self-driven \& personality-like'' ; reduced selection of ``negative traits'' in Phase~2
&
\textbf{Affective carry-over \& reflective continuity}: users report fewer ``reset'' moments and more persona stability across days &
Stabilize a core persona while allowing local flexibility; surface small memory callbacks to make continuity visible. \\
\midrule
Q5 emotional improvement over time, stronger in ``NoEmo→Full'' group (Table~\ref{tab:Q4Q5_within}) &
\textbf{Anticipatory alignment}: agent modulates initiative and tone around upcoming stressors (exams, deadlines) &
Prioritize anticipatory prompts when stress signals accumulate. \\
\midrule
AdI/ESU $>$ BGI for coherence/harmony in experienced users (Table~\ref{tab:ctem-mechanisms}) &
\textbf{Emotional Harmony \& Stable Persona/Tone}: users focused far more on \textit{Auri}’s emotional attunement and persona stability than on its behavior generation &
Invest in online modulation and state transitions; keep behavior pools contextually grounded. \\
\bottomrule
\end{tabular}
\vspace{0.2cm}
\caption{Joint display aligning quantitative effects with qualitative mechanisms and design implications.}
\label{tab:joint-display}
\end{table*}

\section{Discussion}

In this section, we further discuss the design tensions, ethical considerations, broader implications, and limitations.

\subsection{Design Tensions}
Our results suggest that CTEM enhanced participants’ sense of coherence and harmony.
Many participants described \textit{Auri} as continuous across days yet adaptive to context, highlighting the value of CTEM and pointing to directions for future agent design.
Further discussion reveals that this continuous and adaptive experience requires balancing tensions rather than optimizing one factor. 
Accordingly, for \textbf{RQ3} and \textbf{DG3}, we identify four design tensions: \textbf{stability vs.\ variation}, \textbf{coherence vs.\ flexibility}, \textbf{support vs.\ reciprocity}, and \textbf{consistency vs.\ adaptation}. These design tensions are grounded in both quantitative and qualitative findings from our study and can be mapped to corresponding psychological theories (Table~\ref{tab:design-tensions-psychology}).
Taken together, these design tensions synthesize our empirical findings into a coherent explanatory framework of how cross-temporal emotional modeling supports long-term companionship.


\begin{table*}[ht] \centering \renewcommand{\arraystretch}{1.2}\begin{tabular}{>{\raggedright\arraybackslash}p{3.5cm} >{\raggedright\arraybackslash}p{5cm} >{\raggedright\arraybackslash}p{7.5cm}} \toprule \textbf{Design Tension} & \textbf{Psychological Theories} & \textbf{Implications} \\ \midrule Stability vs.\ Variation & Big Five (Conscientiousness vs. Openness); security vs. novelty & Balance routine with novelty to sustain trust and interest. \\ \midrule Coherence vs. Flexibility & Emotion regulation; dynamic models of emotion & Ensure long-term consistency while adapting to context. \\ \midrule Support vs. Reciprocity & Social exchange; attachment; reciprocity norm & Blend support with reciprocity to enhance authenticity. \\ \midrule Consistency vs. Adaptation & Person–situation debate; self-consistency; social cognition & Keep a stable core personality with peripheral flexibility. \\ \bottomrule \end{tabular} \vspace{0.2cm}
\caption{Design tensions with related theories and implications for companion agents.} \label{tab:design-tensions-psychology} \end{table*} 

\textbf{Stability vs.\ Variation} Stability builds trust, but too much stability feels monotonous. Variation adds vitality, but excess unpredictability disrupts continuity. Companionship depends on balance. Interviews revealed divergent views: some users enjoyed diversity as entertainment and looked forward to \textit{Auri}’s next behavior, while others preferred \textit{Auri} to remain simple and stable. 

\recolor{This divergence was reflected in the qualitative analysis as different interpretations of variation. 
On the one hand, many participants placed strong emphasis on ``Stable Persona \& Tone'' and ``Memory Continuity''.
When using the baseline or control system, participants reported clear discomfort with abrupt, insufficiently grounded changes, which they described as ``Breaks in Continuity''.
On the other hand, under the CTEM framework, \textit{Auri} exhibited more complex emotional features and behaviors, such as anticipation, regret, and hesitation. These forms of variation can be interpreted as signals of development within a continuous interaction arc, reflecting some users’ expectations for emotional depth and progression.
One remarked, ``I can fully accept expressions of regret or hesitation from \textit{Auri}. For me, this feels like positive feedback—it makes me feel that \textit{Auri} is taking me seriously.'' 
These divergences indicate that stability underpins trust, whereas moderate variation is expected in interaction, and the two demands mutually constrain each other along a shared interaction trajectory. In other words, building upon a coherent long-term interaction trajectory, measured and smoothly integrated variation can influence user experience. The relationship between long-term stability and localized variation requires careful calibration to enhance engagement without undermining overall coherence.}

Design implication: consequently, agents may adapt the way they generate behaviors to match user tolerance for diversity and need for predictability, leading to more personalized and sustainable companionship. 
In practice, this could be operationalized through a range of algorithmic mechanisms that regulate behavioral variation and stability, enabling agents to adapt this balance in line with individual user preferences.


\textbf{Coherence vs. Flexibility} 
\recolor{Our experimental and interviews highlight users’ demand for agent expressions that are both coherent and flexible.
The layered emotional modeling design proposed in our study effectively supports this demand.}

\recolor{Within the CTEM framework, the ``Behavioral'' and ``Emotional'' modules exert cross-temporal influence. Through sustained behavioral patterns and emotional expression, they jointly shape users’ judgments of \textit{Auri}’s overall coherence. 
Results from Q1 support this design rationale.  Removing either the ``Behavioral'' or ``Emotional'' module led to a substantial decline in attributing coherent traits to \textit{Auri}, indicating that these cross temporal modules play a central role in shaping users’ perception of coherence.}
\recolor{Meanwhile, even in the absence of multimodal output, participants continued to report relatively high level of perceived coherence.
This finding provides preliminary evidence that perceived coherence is not solely driven by surface-level expressive forms; coherence in long-term behavioral patterns and emotional expression can offer users a solid and reliable sense of coherence.}

\recolor{Further experiments and interviews showed that restoring multimodal capabilities led participants to perceive more flexible and richer expressions. One marked, “When \textit{Auri} had access to multimodal elements, its expressions became richer and more flexible, improving the overall experience.”
We interpret these multimodal elements(e.g., emojis, images, social timelines)as enhancing flexibility in short-term emotional expression, particularly at the expressive and visual levels, thereby further strengthening users’ perception of coherence.}
This suggests that, within layered emotional design, cross-temporal emotional and behavioral expressions play a critical role in users’ experience of coherence, while multimodal cues enrich expressive flexibility in short-term interactions.

An open question remains: should agents display authentic human-like emotions, or more masked expressions like those often used in instant messaging? Future research should examine which path better supports user experience. 

\recolor{To enhance agent coherence, this study goes beyond emotional modeling by examining the role of maintaining stable value goals and personality core.
Q2 compared the relative importance of different positive traits in shaping perceived coherence. Results indicate that ``Self-driven \& personality-like''
 traits were viewed as the most critical, followed by ``Relational consistency'' and ``Social behavior''.
This ranking remained stable over time, suggesting that users developed a consistent understanding of \textit{Auri}’s coherence. Participants tended to attribute coherence to \textit{Auri}’s stable value goals and personality core, rather than to a simple accumulation of momentary behaviors or fragmented states.
These value goals and personality core continuously shape the agent’s behavioral tendencies and emotional dynamics and, together with short-term contextual factors, guide state updates and interaction behaviors. This enables the agent to remain coherent and predictable over time while retaining short-term contextual flexibility.}

In sum, coherence and flexibility operate at different temporal scales along the same interaction trajectory and are better understood as a balance rather than independent optimization goals.

Design Implication: by maintaining and adhering to its distinctive value goals, an agent can present users with characteristic patterns of both coherence and flexibility. In practice, combining layered emotional models with value-goal computation models may help agents flexibly adapt to evolving contexts while still maintaining overall coherence.

\textbf{Support vs. Reciprocity} 
\recolor{In our study, \textit{Auri}’s ``Emotional'' module primarily functioned to enhance supportive interaction. Results in Q4 and Q5 indicate that restoring the ``Emotional'' module improved users’ perceived contextual appropriateness and led to emotional improvement. 
When participants compared the baseline system with the full CTEM system, these improvements became more pronounced and were more often attributed to mechanisms related to Adaptive Interaction (AdI) and Emotional State Updating (ESU).}

\recolor{Beyond supportive responses, \textit{Auri} occasionally expressed frustration, self-mockery, mild complaints, or refusal. Small signs of negative emotion conveyed vulnerability and increased perceived authenticity. }One remarked, “seeing \textit{Auri} so unlucky made me feel my own life was not so bad.” This suggests that both positive and negative emotions shape experience: positive emotions bring comfort, while negative ones can also bring relief through comparison.

\recolor{Findings from Q6 further showed that participants tended to engage \textit{Auri} as a lightweight companion embedded in specific daily moments, rather than as a source of deep emotional support.}
We argue that limited reciprocal expressions, such as occasional displays of need or vulnerability, could create a sense of mutual care and made users feel needed. However, without regulation, frequent negative expressions or excessive reciprocal demands might risk causing emotional fatigue.

In sum, within the same companion role, supportive and reciprocal responses serve different emotional expectations and do not always scale together in enhancing emotional support.

Design implication: designers should consider how to balance supportive and reciprocal dynamics to foster authenticity without causing burden. 

\textbf{Consistency vs. Adaptation} Gradual personalization deepened engagement, but over-accommodation threatened consistency.
\recolor{In the Q3 investigation of \textit{Auri}’s adjust ability, many participants customized \textit{Auri}’s personality traits, often making it more distinctive or even aggressive. Participants frequently interpreted such adjust ability as increasing perceived fit and flexibility, and in this context it was generally not viewed as undermining coherence.  
But several later reverted to the original state.} This pattern reflects an initial desire for novelty followed by a return to stability and control. 
Design implication: personality customization should preserve a persistent core while allowing peripheral flexibility.

\subsection{Ethical Considerations and Safeguards}

Our research focuses on the safe interaction between young people and empathetic agents. In the design of companion agents with cross-temporal emotional modeling, ethical and risk-related concerns are inevitable. While this work has emphasized avoiding romanticization and over-dependence, it is crucial to systematically address potential issues related to long-term reliance, data privacy, and mental health risks. 

\textbf{Long-term reliance.} With the introduction of cross-temporal memory, users may gradually develop excessive dependence on the agent, potentially weakening their real-world social connections. Although CTEM enhances engagement by coupling affect and behavior across time, this temporal coherence may also intensify users’ anthropomorphic perceptions of the agent, thereby increasing the risk of emotional over-attachment. Recent longitudinal evidence further indicates that higher levels of anthropomorphism can mediate the social impacts of companion chatbots, with users who anthropomorphize agents more strongly showing greater shifts in human–human social outcomes \cite{guingrich2025longitudinal}. \mycolor{To mitigate these risks, our design follows a principle of lightweight companionship: (1) Visual identity. \textit{Auri}’s visual design intentionally diverges from human-like representations; it adopts a soft, pet-inspired appearance that conveys warmth and empathy while subtly reminding users of its non-human nature. (2) Behavioral design. The behavioral generation of \textit{Auri} emphasizes characteristics of a young, pet-like companion rather than striving to imitate human behavior or emulate realistic human agency. (3) Real-world grounding. The system incorporates real-world time cues and holiday events to encourage authentic offline social interactions during meaningful moments, thereby preventing excessive immersion in virtual companionship.}

\textbf{Data privacy.} CTEM does not record users’ raw emotional trajectories. Instead, the system adheres to a principle of minimal storage, retaining only abstract representations relevant to emotional modeling. 

\textbf{Mental health risks.} When users feel emotionally distressed or express psychological concerns, there is a risk that the agent may be misperceived as a substitute for professional counseling. To avoid overstepping, the system explicitly frames itself as a \textit{companion assistant} rather than a \textit{therapist}. Upon detecting signals of risk, the agent employs conversational de-escalation, shifting toward neutral reassurance, and will proactively ask whether the user wishes to be connected to professional counseling services or school/university mental health staff. This ensures that support is redirected to appropriate professional channels when necessary.

\textbf{Age-group selection.} The primary motivation of this study is to understand how humans interact with intelligent agents in emotional companionship scenarios and to examine how cross-temporal emotional cues shape users’ affective experiences and behavioral responses. To address this goal, we initially recruited participants aged 18–26. This age group exhibits relatively consistent patterns in socioemotional goals, emotional expression styles, and interaction preferences, for example, a stronger tendency to seek emotional feedback, more direct reactions to affective cues, and higher frequencies of peer-oriented social interactions. These characteristics make their behavioral responses in emotional companionship systems more observable and easier to quantify ~\cite{zhaoyang2018}. In addition, young adults generally possess higher levels of digital literacy and familiarity with interactive technologies, enabling more natural engagement with the system and reducing noise in experimental observations ~\cite{venkatesh2003}. \mycolor{However, the developmental specificity of this age group limits the generalizability of our findings to older populations, whose emotional needs and interaction motivations often place greater emphasis on relational stability and deeper emotional meaning ~\cite{carstensen1999}. Future work should therefore include a broader range of age groups to investigate how different forms of emotional companionship manifest across the human lifespan.}

Based on these reflections, we propose a set of ethical design principles: 

\begin{itemize}
  \item \textbf{Safeguarded Companionship}:\\
  \mycolor{Maintain lightweight and bounded companionship to prevent emotional over-attachment that may arise from prolonged or anthropomorphized interactions.}

  \item \textbf{Transparency \& Control}:\\
  Ensure users are informed about, and can manage, stored data and memories.

  \item \textbf{Responsible Withdrawal}:\\
  Employ conversational de-escalation and withdrawal strategies in risk scenarios.

  \item \textbf{Bounded Roles}:\\
  Clearly define the agent’s role boundaries to avoid confusion with mental health services.

  \item \textbf{Wellbeing Nudges}:\\
  Encourage real-world social interactions by incorporating time- and holiday-based reminders.
\end{itemize}

These principles not only help mitigate potential risks but also provide actionable guidance for the responsible design of future companion agents.

\subsection{Broader Implications}

By positioning agents as companions rather than romantic partners, CTEM offers a safer trajectory for long-term adoption. Such designs can promote wellbeing across healthcare, education, and community contexts, where sustained but bounded support is valuable. Commercial precedents illustrate divergent directions; CTEM advances the companion model by integrating continuity with harmony, offering a balanced alternative that prioritizes wellbeing and sustainable human–AI interaction.

\subsection{Limitations and Future Work}

This study presents an initial exploration through a 21-day deployment with 96 participants (18–26 years old). 
The relatively short study duration, unequal exposure durations in sub-study, and focused demographic limit generalizability; longer-term and more balanced studies with more diverse cohorts are needed.
While our joint display clarifies several CTEM pathways, causal attribution remains tentative; future work should embed \textit{planned} qualitative probes (e.g., diary-elicited, event-contingent interviews) synchronized with \textit{a priori} quantitative contrasts to strengthen inference. Extending modalities such as voice, strengthening comparisons with stronger baselines (memory/persona-enabled agents), and formalizing safeguards against over-dependence \mycolor{(boundary cues/lightweight companionship strategies)} are further priorities.

\section{Conclusion}
This work introduced Cross-Temporal Emotional Modeling (CTEM), a framework that integrates long-term behavioral histories with moment-to-moment emotional expressions to support more coherent and responsive companion AI. Through a 21-day deployment with 96 participants, we identified four key design tensions: stability versus variation, coherence versus flexibility, support versus reciprocity, and consistency versus adaptation. Each is grounded in psychological theory and manifested in everyday interactions, providing actionable guidance for future systems. While our study was limited in scale and duration, it serves as an initial empirical demonstration of how cross-temporal modeling can enrich daily companionship experiences. Future work should extend CTEM to multimodal contexts, larger and more diverse populations, and longer-term deployments, while also engaging more deeply with the ethical dimensions of emotionally adaptive companions. Taken together, this work establishes a theoretical and practical foundation for developing AI systems that are not only intelligent but also emotionally durable in sustained use.

\begin{acks}
This work was supported by the ``Fundamental Research Funds for the Central Universities'' (Grant No. CUC24SG014).
\end{acks}

\bibliographystyle{ACM-Reference-Format}
\bibliography{reference}

\appendix
 \renewcommand\thesection{A\arabic{section}}
 \setcounter{table}{0}
 \renewcommand{\thetable}{A\arabic{table}}
 \setcounter{figure}{0}
 \renewcommand{\thefigure}{A\arabic{figure}}
 \setcounter{algorithm}{0}
 \renewcommand{\thealgorithm}{A\arabic{algorithm}}

\section{Appendix}

\subsection{Algorithm Flow and Prompts}

\subsubsection{\textbf{CTEM Status Representation Details}}
The CTEM status representation formalizes the agent’s dynamic internal state and its 
interaction with behavior selection. It unifies physical vitality, affective dimensions, 
motivational drives, and structural components into a coherent schema. As shown in 
Table~\ref{tab:ctem-schema}, three layers of information are specified:

\begin{table}[ht]
\centering
\tiny
\renewcommand{\arraystretch}{1.0}
\setlength{\tabcolsep}{4pt}
\begin{tabular}{
  >{\raggedright\arraybackslash}p{2.4cm}
  >{\raggedright\arraybackslash}p{3.0cm}
  >{\raggedright\arraybackslash}p{1.0cm}
  >{\raggedright\arraybackslash}p{0.9cm}
}
\toprule
\textbf{Variable} & \textbf{Description} & \textbf{Range / Format} & \textbf{Default} \\
\midrule
\texttt{bio\_value} & Current level of physical energy (vitality) & $[0,1]$ & 0.5 \\
\texttt{psycho\_valence\_value} & Current emotional valence & $[-1,1]$ & 0.0 \\
\texttt{psycho\_arousal\_value} & Current arousal level & $[0,1]$ & 0.5 \\
\texttt{bio\_require} & Minimum physical energy required & $[0,1]$ & 0.1 \\
\texttt{bio\_consumption} & Change in physical energy after execution & $[-1,1]$ & -0.1 \\
\texttt{user\_familiarity} & Degree of social familiarity with the user & $[0,1]$ & 0.0 \\
\addlinespace[2pt]

\texttt{bio\_physiological\_drive} & Instinctive needs (bodily vitality drive) & $[0,1]$ & 0.5 \\
\texttt{bio\_pain\_avoidance} & Harm avoidance (risk/injury avoidance) & $[0,1]$ & 0.5 \\
\texttt{bio\_health\_preservation} & Health preservation (minimizing depletion) & $[0,1]$ & 0.5 \\
\addlinespace[2pt]

\texttt{psycho\_emotional\_reactivity} & Emotional reactivity (event-triggered fluctuation) & $[0,1]$ & 0.5 \\
\texttt{psycho\_risk\_aversion} & Risk aversion (conflict/uncertainty avoidance) & $[0,1]$ & 0.5 \\
\texttt{psycho\_goal\_persistence} & Goal persistence (long-term goals) & $[0,1]$ & 0.5 \\
\texttt{psycho\_curiosity\_drive} & Curiosity drive (novelty seeking) & $[0,1]$ & 0.5 \\
\addlinespace[2pt]

\texttt{social\_norm\_adherence} & Norm adherence (following rules) & $[0,1]$ & 0.5 \\
\texttt{social\_prosocial\_motivation} & Prosocial motivation (helping others) & $[0,1]$ & 0.5 \\
\texttt{social\_self\_presentation} & Self-presentation (maintaining image/dignity) & $[0,1]$ & 0.5 \\
\texttt{social\_role\_duty\_sense} & Role duty sense (role-based obligations) & $[0,1]$ & 0.5 \\
\texttt{social\_group\_affiliation} & Group affiliation (loyalty/collective goals) & $[0,1]$ & 0.5 \\
\addlinespace[2pt]

$M_t$ & Memory (self, user, context) & dynamic store & empty \\
$G_t$ & Personality prompt modulated by $H_t$ & text prompt & baseline $G$ \\
$B$ & Behavior pool (5 categories) & structured list & predefined \\
$B_t$ & Behavioral inventory $\langle Past, Present, Future \rangle$ & dynamic struct. & empty \\
$e_t$ & Daily episodic summary & text summary & none \\
$\mathcal{G}_{safe}$ & Safe personality space & constraint set & predefined \\
\bottomrule
\end{tabular}
\captionsetup{width=0.95\linewidth}
\caption{Comprehensive schema of CTEM variables, including state values, behavior requirements, motivational dimensions, structural components, and user familiarity.}
\label{tab:ctem-schema}
\end{table}

\subsubsection{\textbf{Auri Personality Configuration Examples}}

To demonstrate how Auri can be flexibly configured, we present three illustrative personality profiles(Tables~\ref{personality1}--\ref{personality3}). 
Each profile highlights different tendencies across the biological, psychological, and social dimensions, 
showing how diverse behavioral traits can emerge from different parameterizations. 

\begin{table}[ht]
\centering
\tiny
\renewcommand{\arraystretch}{1}
\setlength{\tabcolsep}{4pt}
\begin{tabular}{p{3.5cm} p{3.6cm}}
\toprule
\textbf{Trait} & \textbf{Value / Interpretation} \\
\midrule
bio\_physiological\_drive & 0.35 (moderate vitality) \\
bio\_pain\_avoidance & 0.25 (low avoidance) \\
bio\_health\_preservation & 0.4 (moderately high) \\
psycho\_emotional\_reactivity & 0.3 (moderate) \\
psycho\_risk\_aversion & 0.3 (low) \\
psycho\_goal\_persistence & 1.0 (very high $\rightarrow$ strong learner) \\
psycho\_curiosity\_drive & 1.0 (moderately high) \\
social\_norm\_adherence & 0.15 (very low) \\
social\_prosocial\_motivation & 0.06 (very low) \\
social\_self\_presentation & 0.06 (very low) \\
social\_role\_duty\_sense & 0.06 (very low) \\
social\_group\_affiliation & 0.06 (very low) \\
\bottomrule
\end{tabular}
\caption{Personality 1: A learner with low social\_needs.}
\label{personality1}
\end{table}

\begin{table}[ht]
\centering
\tiny
\renewcommand{\arraystretch}{1}
\setlength{\tabcolsep}{4pt}
\begin{tabular}{p{3.5cm} p{3.6cm}}
\toprule
\textbf{Trait} & \textbf{Value / Interpretation} \\
\midrule
bio\_physiological\_drive & 0.4 (moderately high energy) \\
bio\_pain\_avoidance & 0.35 (moderate) \\
bio\_health\_preservation & 0.4 (moderate) \\
psycho\_emotional\_reactivity & 0.5 (high) \\
psycho\_risk\_aversion & 0.4 (moderately high) \\
psycho\_goal\_persistence & 0.5 (moderately high) \\
psycho\_curiosity\_drive & 0.4 (moderate) \\
social\_norm\_adherence & 0.5 (high) \\
social\_prosocial\_motivation & 1.0 (very high $\rightarrow$ empathetic) \\
social\_self\_presentation & 1.0 (high, seeks recognition) \\
social\_role\_duty\_sense & 1.0 (very high $\rightarrow$ strong sense of duty) \\
social\_group\_affiliation & 1.0 (high $\rightarrow$ strong social\_needs) \\
\bottomrule
\end{tabular}
\caption{Personality 2: Highly social and empathetic.}
\label{personality2}
\end{table}

\begin{table}[ht]
\centering
\tiny
\renewcommand{\arraystretch}{1}
\setlength{\tabcolsep}{4pt}
\begin{tabular}{p{3.5cm} p{3.6cm}}
\toprule
\textbf{Trait} & \textbf{Value / Interpretation} \\
\midrule
bio\_physiological\_drive & 0.7 (high vitality $\rightarrow$ active lifestyle) \\
bio\_pain\_avoidance & 0.25 (low $\rightarrow$ accepts physical challenge) \\
bio\_health\_preservation & 0.5 (moderate) \\
psycho\_emotional\_reactivity & 0.7 (high $\rightarrow$ cheerful) \\
psycho\_risk\_aversion & 0.2 (low $\rightarrow$ adventurous) \\
psycho\_goal\_persistence & 1.0 (high $\rightarrow$ motivated learner) \\
psycho\_curiosity\_drive & 1.0 (high curiosity) \\
social\_norm\_adherence & 0.6 (moderately high) \\
social\_prosocial\_motivation & 0.8 (high $\rightarrow$ extroverted) \\
social\_self\_presentation & 0.8 (high $\rightarrow$ expressive) \\
social\_role\_duty\_sense & 0.6 (moderately high) \\
social\_group\_affiliation & 0.8 (high $\rightarrow$ enjoys group activities) \\
\bottomrule
\end{tabular}
\caption{Personality 3: Cheerful, energetic learner who enjoys sports.}
\label{personality3}
\end{table}

\subsubsection{\textbf{Behaviors Pool Data Samples}}
\begin{figure}[ht]
    \centering
    \includegraphics[width=1.0\linewidth]{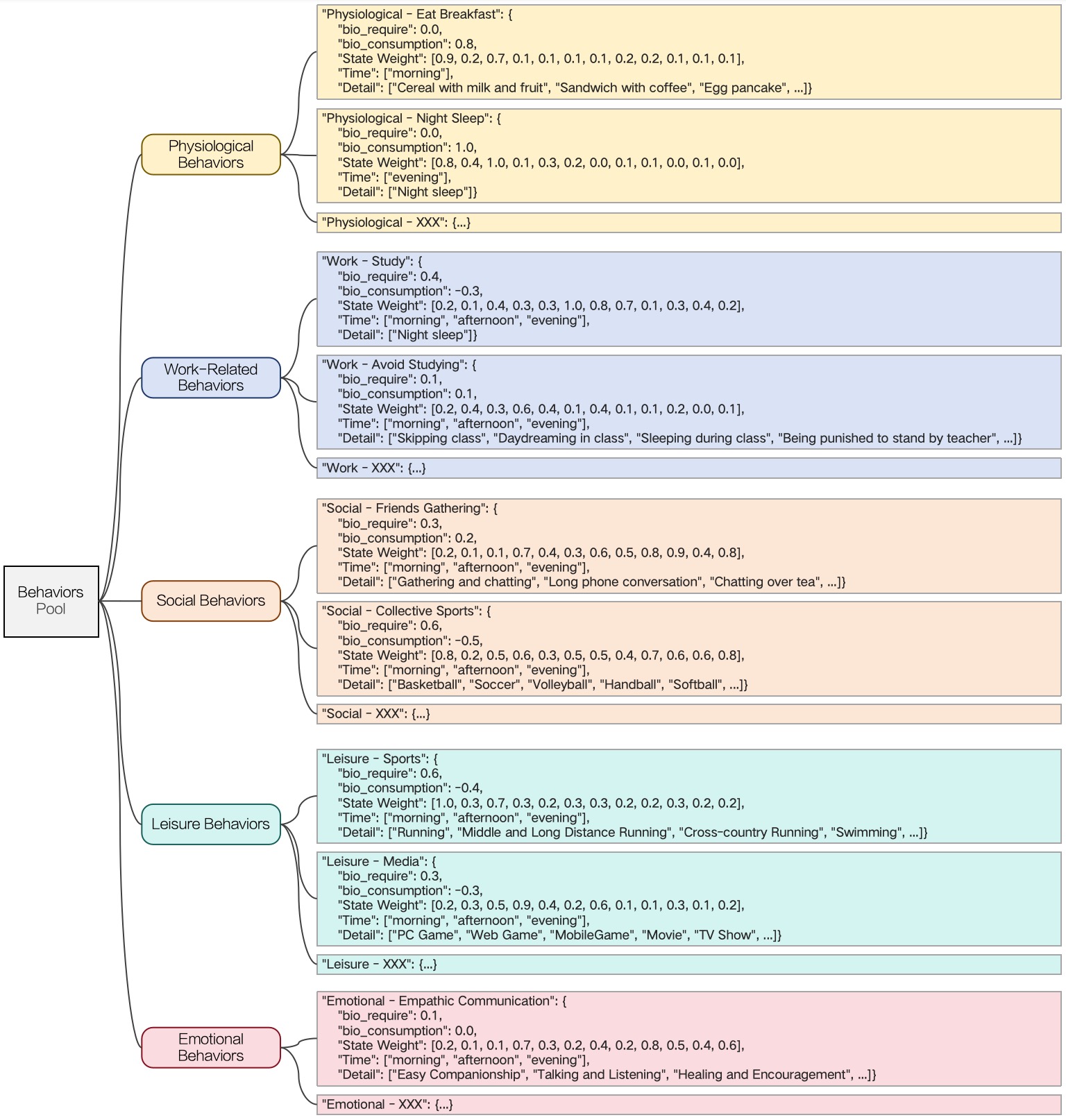}
    \Description{}
    \caption{Behaviors Pool Data Samples.}
    \label{fig:Pool}
\end{figure}

The behaviors pool in CTEM groups agent actions into categories such as physiological, 
work, leisure, social, and emotional. Each action is annotated with energy requirements, 
psycho\_impact, and motivational weights across twelve drives. This structure 
supports flexible and context-aware behaviors selection. Figure~\ref{fig:Pool} shows 
representative samples.

\subsubsection{\textbf{CTEM Flowchart}}

\begin{figure}[ht]
    \centering
    \includegraphics[width=1.0\linewidth]{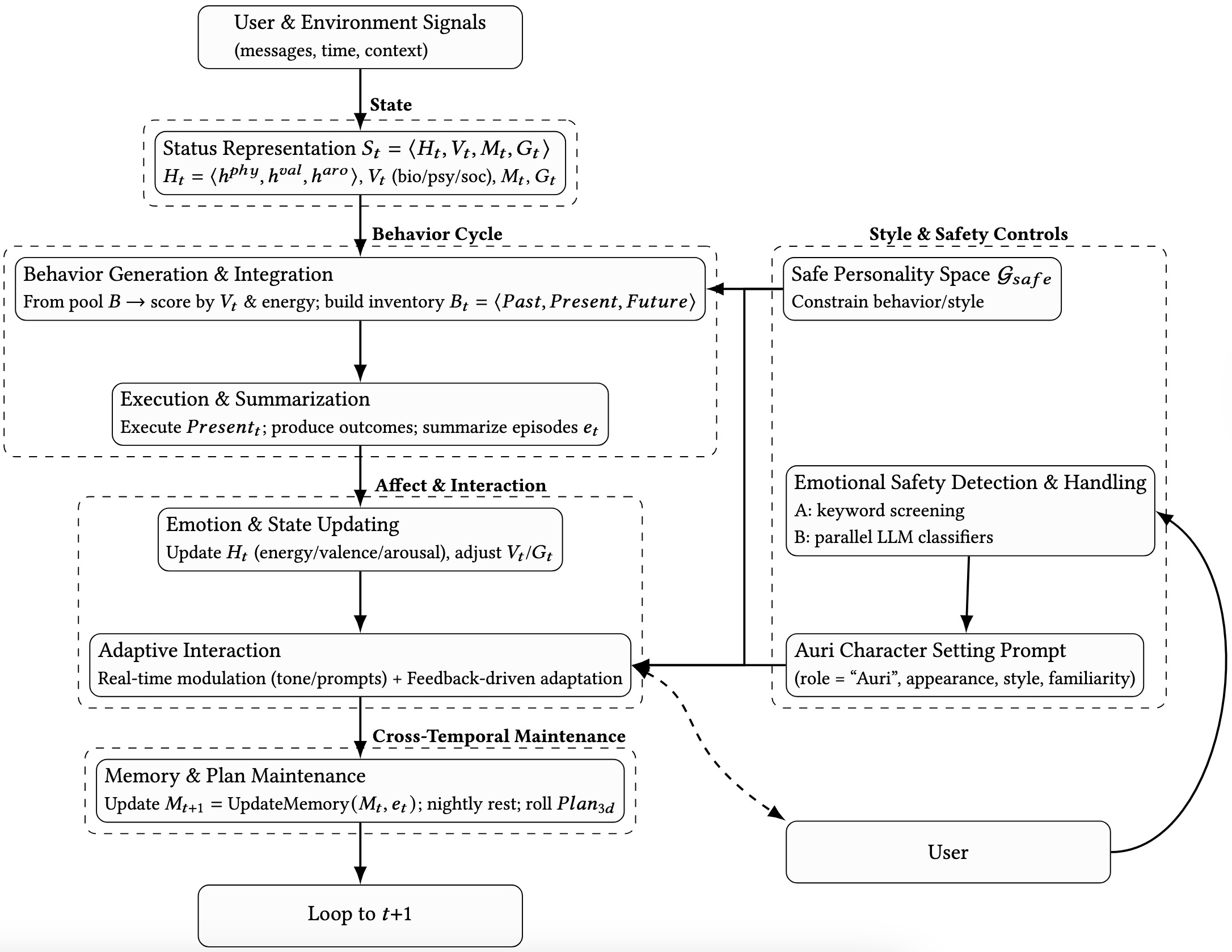}
    \Description{}
    \caption{CTEM high-level flow}
    \label{fig:ctem-highlevel}
\end{figure}

Fig.~\ref{fig:ctem-highlevel} illustrates the high-level workflow of the 
Cross-Temporal Emotional Modeling (CTEM) framework. The process begins with 
user and environmental signals, which are encoded into the agent’s state 
representation. Based on this state, behaviors are generated and integrated 
into a structured inventory that balances past experiences, current conditions, 
and future intentions. Executed behaviors update emotional and motivational 
states, which in turn guide adaptive interaction with the user. 

To ensure safety and personality consistency, all outputs are constrained by a 
safe personality space and modulated by Auri’s character-setting prompt. Memory 
is updated iteratively, supporting long-term adaptation and maintaining 
cross-temporal coherence across sessions.

\subsubsection{\textbf{Auri's Daily Conversation Snapshot}}

\begin{figure}[ht]
  \centering
  \begin{minipage}[t]{0.32\linewidth}
    \centering
    \includegraphics[width=\linewidth]{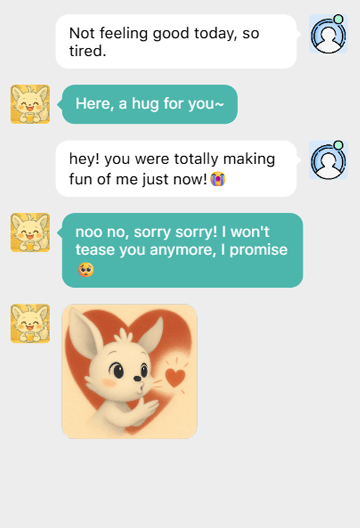}
  \end{minipage}\hfill
  \begin{minipage}[t]{0.32\linewidth}
    \centering
    \includegraphics[width=\linewidth]{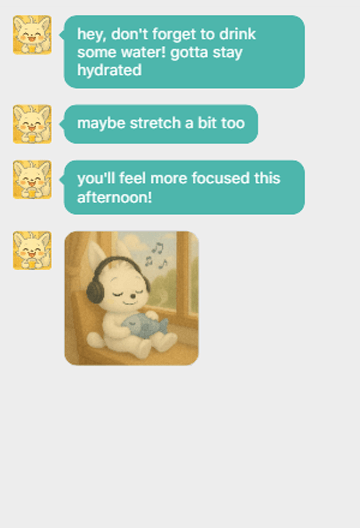}
  \end{minipage}\hfill
  \begin{minipage}[t]{0.32\linewidth}
    \centering
    \includegraphics[width=\linewidth]{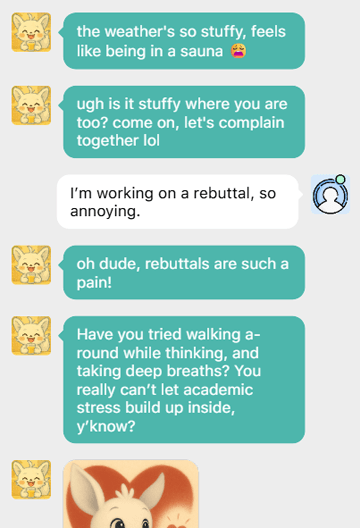}
  \end{minipage}
  \Description{}
  \caption{Examples 1. Auri proactively provided comfort to the participants.}
  \label{fig:dialog-multimode}
\end{figure}

\begin{figure}[ht]
  \begin{minipage}[t]{0.32\linewidth}
    \centering
    \includegraphics[width=\linewidth]{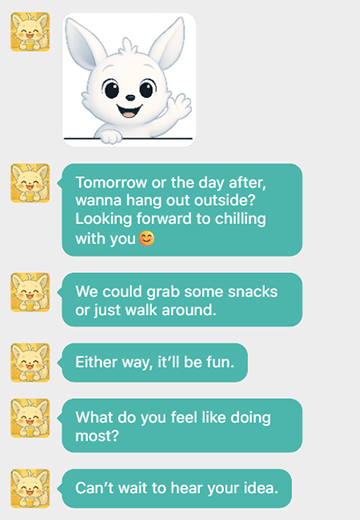}
  \end{minipage}\hfill
  \begin{minipage}[t]{0.32\linewidth}
    \centering
    \includegraphics[width=\linewidth]{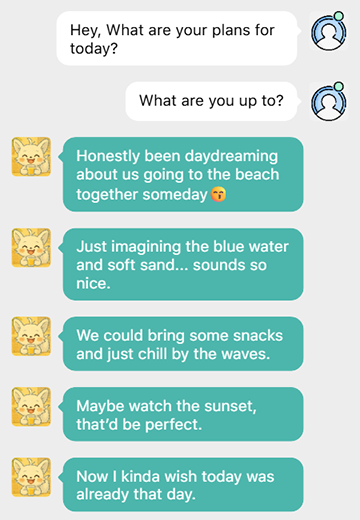}
  \end{minipage}\hfill
  \begin{minipage}[t]{0.32\linewidth}
    \centering
    \includegraphics[width=\linewidth]{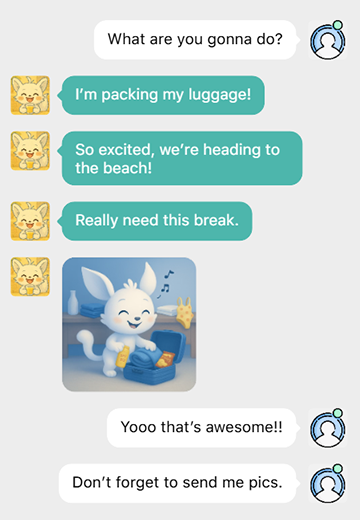}
  \end{minipage}
  \Description{}

  \caption{Examples 2. Auri engaged the participants in conversations about their future plans.}
  \label{fig:dialog-multimode2}
\end{figure}

\begin{figure}[ht]
 \begin{minipage}[t]{0.32\linewidth}
   \centering
   \includegraphics[width=\linewidth]{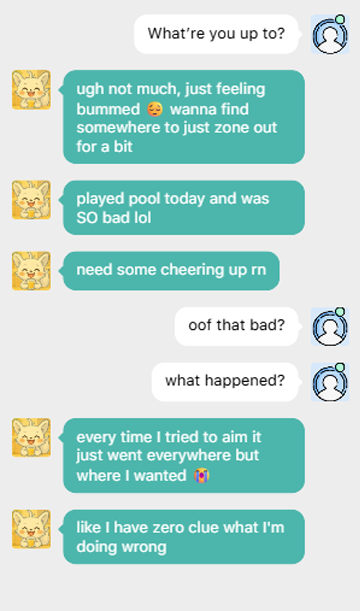}
 \end{minipage}\hfill
 \begin{minipage}[t]{0.32\linewidth}
   \centering
   \includegraphics[width=\linewidth]{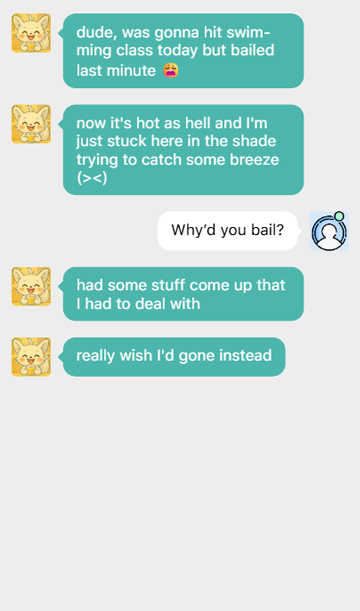}
 \end{minipage}\hfill
 \begin{minipage}[t]{0.32\linewidth}
   \centering
   \includegraphics[width=\linewidth]{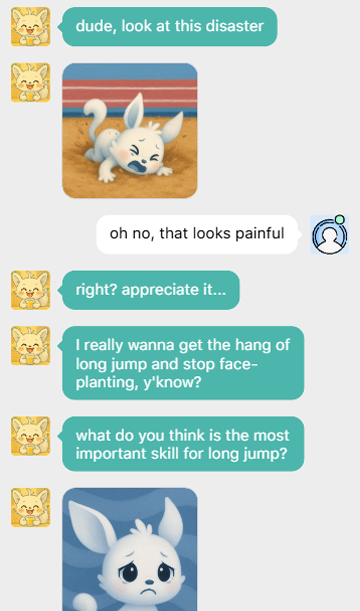}
 \end{minipage}
 \Description{}
 \caption{Examples 3. Auri expressed grievances to the participants regarding its unfortunate encounters.}
 \label{fig:dialog-multimode3}
\end{figure}

\begin{figure}[ht]
 \begin{minipage}[t]{0.32\linewidth}
   \centering
   \includegraphics[width=\linewidth]{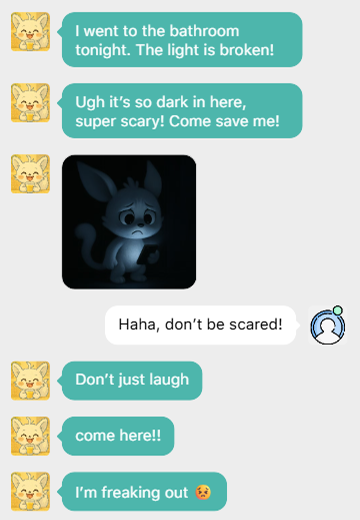}
 \end{minipage}\hfill
 \begin{minipage}[t]{0.32\linewidth}
   \centering
   \includegraphics[width=\linewidth]{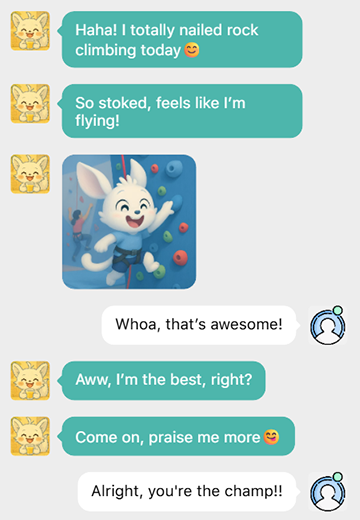}
 \end{minipage}\hfill
 \begin{minipage}[t]{0.32\linewidth}
   \centering
   \includegraphics[width=\linewidth]{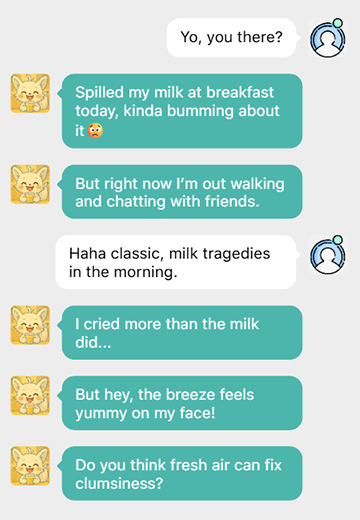}
 \end{minipage}
 \Description{}
 \caption{Examples 4. Auri shared its emotional states with the participants.}
 \label{fig:dialog-multimode4}
\end{figure}

\begin{figure}[ht]
 \begin{minipage}[t]{0.30\linewidth}
   \centering
   \includegraphics[width=\linewidth]{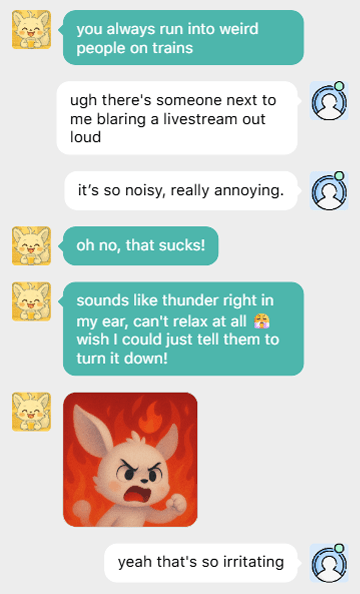}
 \end{minipage}\hfill
 \begin{minipage}[t]{0.30\linewidth}
   \centering
   \includegraphics[width=\linewidth]{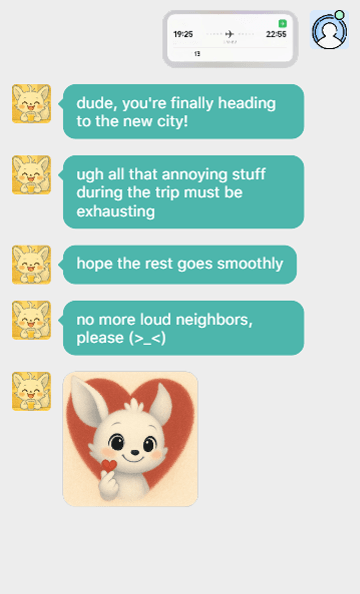}
 \end{minipage}\hfill
 \begin{minipage}[t]{0.30\linewidth}
   \centering
   \includegraphics[width=\linewidth]{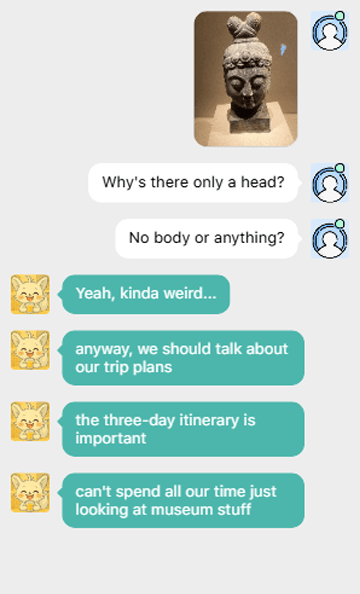}
 \end{minipage}
 \Description{}

 \caption{Examples 5. The participant, during his travels, talked with Auri about the journey.}
 \label{fig:dialog-multimode5}
\end{figure}

Figures ~\ref{fig:dialog-multimode} –~\ref{fig:dialog-multimode5} illustrate the daily conversations between Auri and the user. They showcase different dialog scenarios and emotional expressions in everyday interactions.

\subsubsection{Emojis and Stickers}
\begin{figure}[ht]
  \centering
  \includegraphics[width=0.9\linewidth]{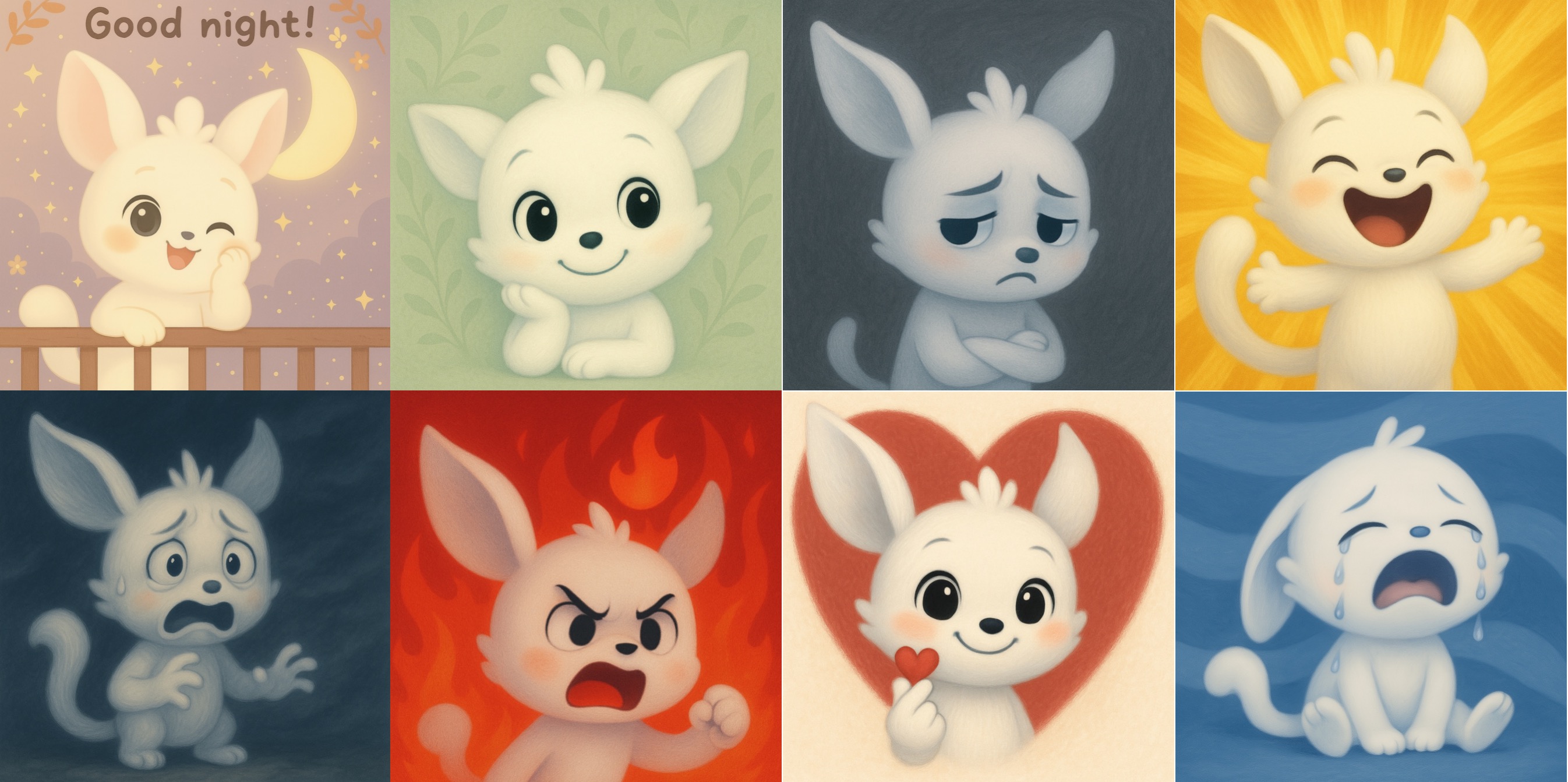}
\captionsetup{width=0.95\linewidth}
\Description{}

  \caption{Emoji meanings. Row 1 (left to right): good night, companionship, emo, happy. Row 2 (left to right): scared, angry, finger heart, sad.}
  \label{fig:emoji}
\end{figure}

To enhance the effectiveness of emotional expression, we designed a comprehensive emoji set that the LLM dynamically selects and delivers based on the emotions inferred from the conversation content (Fig.~\ref{fig:emoji}).

\subsubsection{Auri social\_Timeline}
\begin{figure}[ht]
\begin{minipage}[t]{0.5\linewidth}
  \centering
   \includegraphics[width=\linewidth]{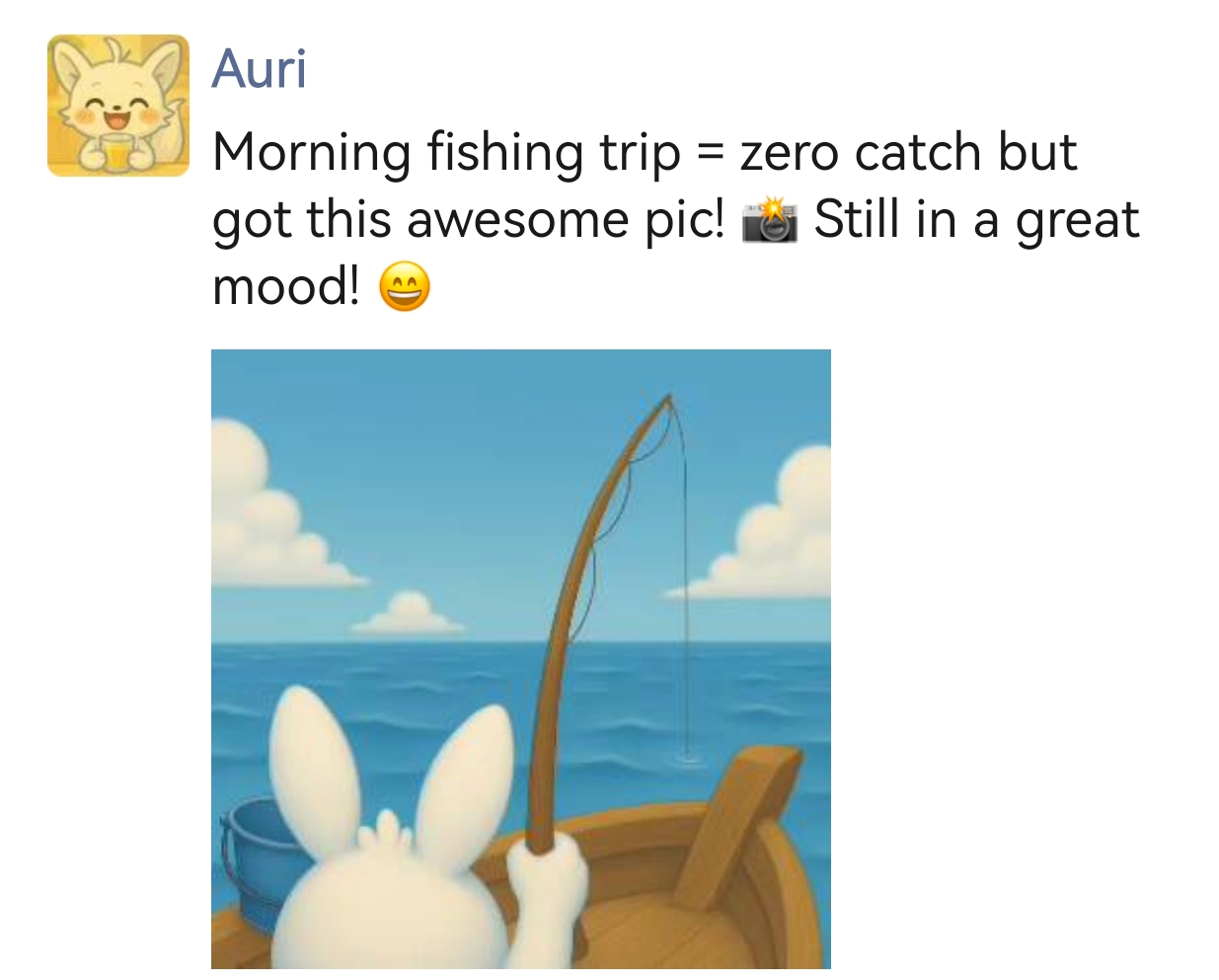}
\end{minipage}\hfill
\begin{minipage}[t]{0.5\linewidth}
  \centering
  \includegraphics[width=\linewidth]{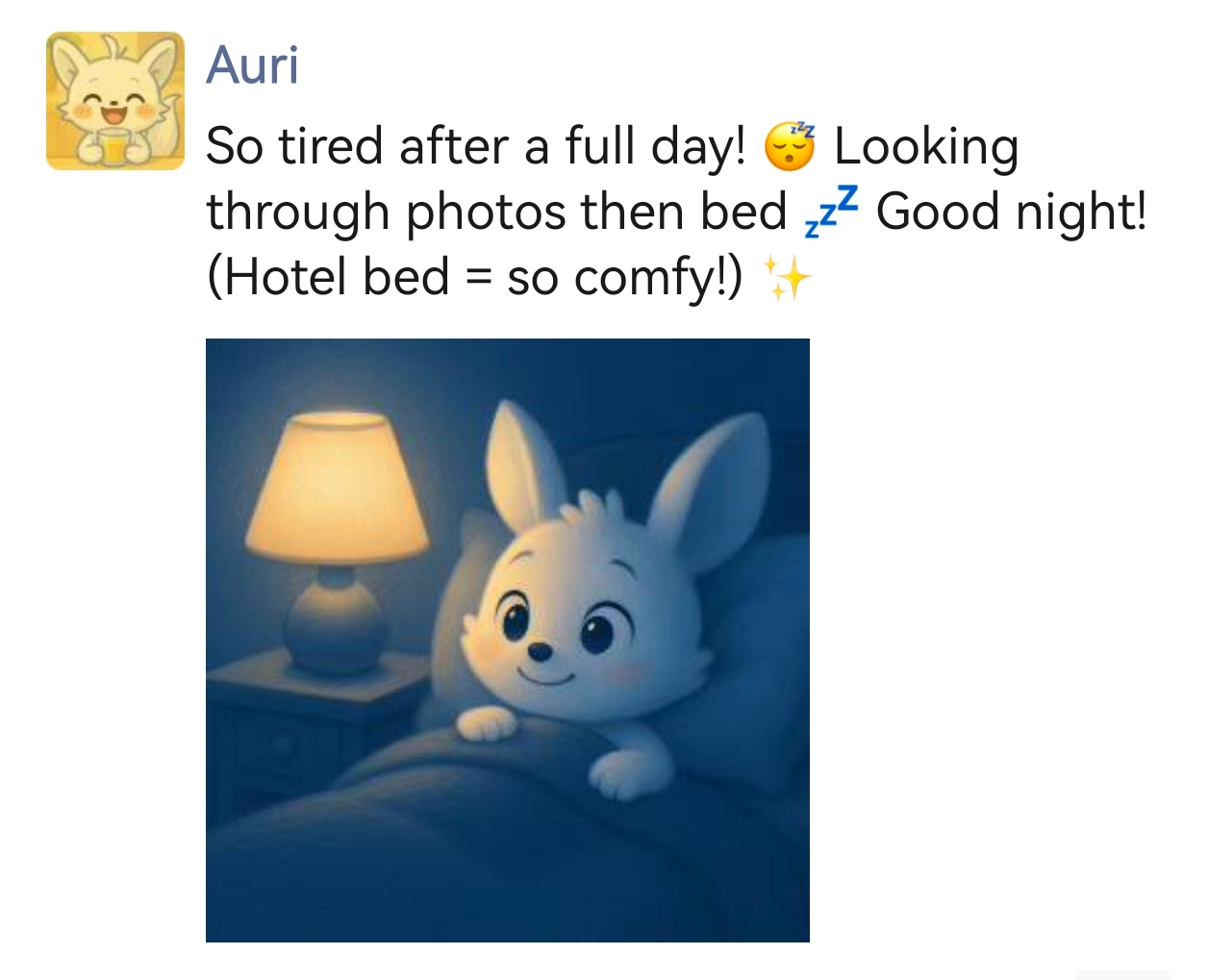}
\end{minipage}\hfill
\Description{}

\caption{Auri’s social timeline (temporal moments).}
\label{fig:social1}
\end{figure}

\begin{figure}[ht]
\begin{minipage}[t]{0.5\linewidth}
  \centering
  \includegraphics[width=\linewidth]{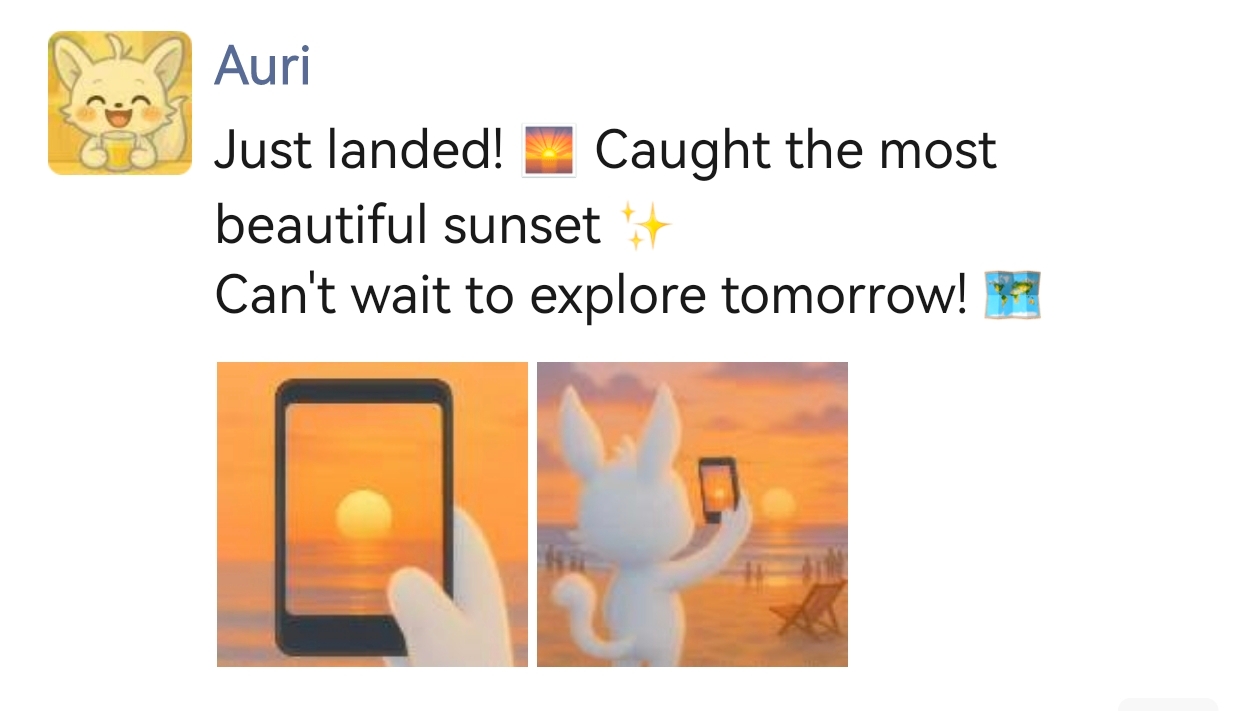}
  \label{fig:social3}
\end{minipage}\hfill
\begin{minipage}[t]{0.5\linewidth}
  \centering
  \includegraphics[width=\linewidth]{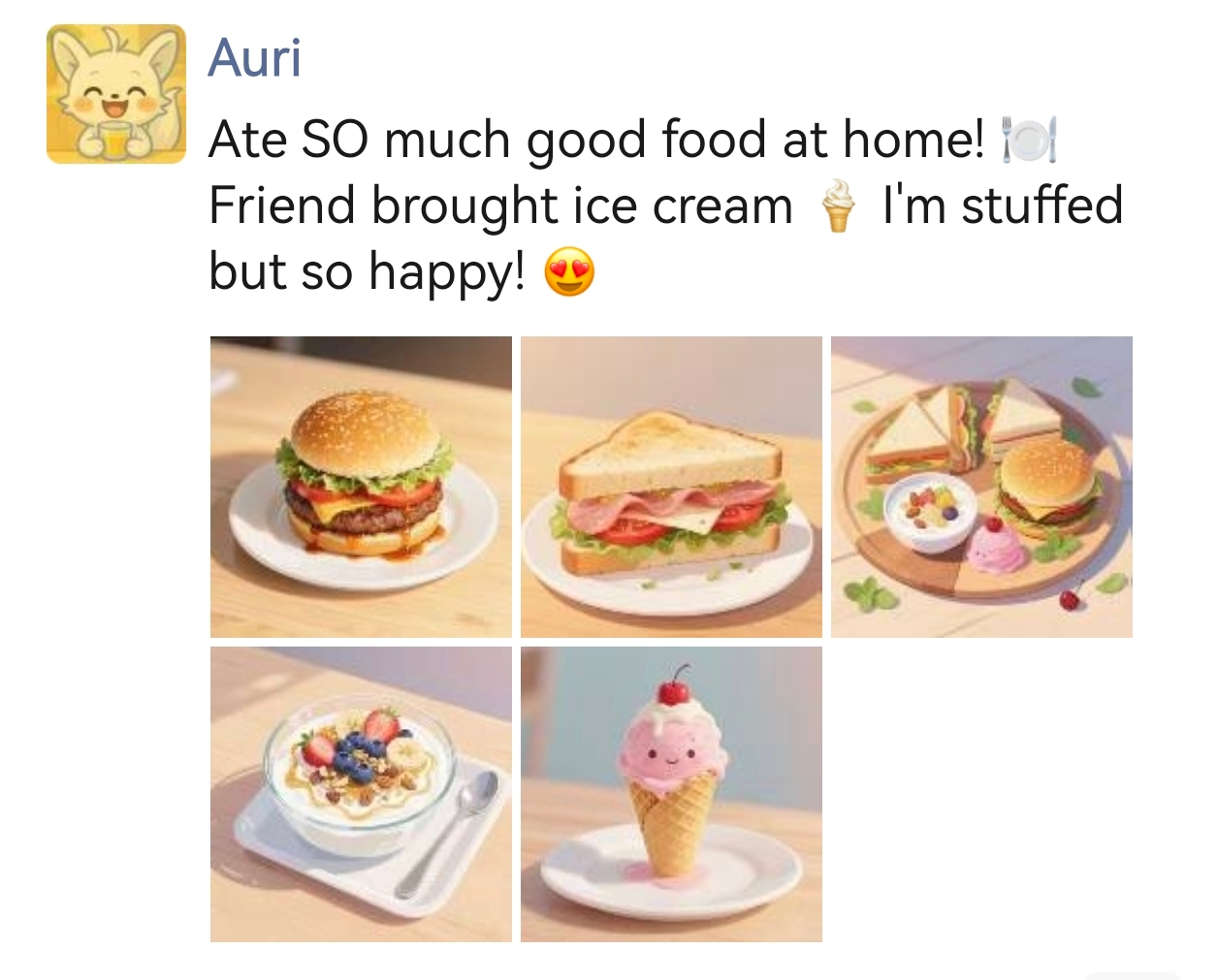}
\end{minipage}\hfill
 \Description{}
 \caption{Auri`s social timeline (everyday activities).}
 \label{fig:social2}
\end{figure}

Auri is system-driven to generate simulated daily activities resembling posts in a social\_timeline(Fig.~\ref{fig:social1}, Fig.~\ref{fig:social2}).

\subsection{Additional Study Design and Results}

\subsubsection{\textbf{Coherent-Trait Perception (Q1)}}
Table~\ref{tab:appendix_coherent_traits} summarizes the marginal probabilities and within-group phase contrasts for coherent-trait perception (\textbf{Q1}), derived from the binomial GEE models reported in the main text.

\begin{table}[htbp]
\centering
\tiny
\begin{tabular}{llcccc}
\toprule
\multirow{2}{*}{\textbf{Group}} & \multirow{2}{*}{\textbf{Trait}} 
& \multicolumn{2}{c}{\textbf{Phase 1}} 
& \multicolumn{2}{c}{\textbf{Phase 2}} \\
\cmidrule(lr){3-4} \cmidrule(lr){5-6}
& & $p$ & 95\% CI & $p$ & 95\% CI \\

\midrule
\multirow{5}{*}{\shortstack[l]{NoMultiModal\\→Full}}
& Self-driven\&personality-like & .833 & [.523, .958] & .750 & [.448, .917] \\
& social\_behavior & .750 & [.448, .917] & .500 & [.244, .756] \\
& Relational consistency & .833 & [.523, .958] & .750 & [.448, .917] \\
& Appearance & .250 & [.083, .552] & .417 & [.185, .692] \\
& No coherent trait & .083 & [.012, .413] & .000 & [.000, .000] \\
\cmidrule{2-6}
& Positive & .917 & [.587, .999] & .999 & [.999, .999] \\

\midrule
\multirow{5}{*}{\shortstack[l]{NoBehav\\→Full}}
& Self-driven\&personality-like & .467 & [.241, .707] & .652 & [.189, .938] \\
& social\_behavior & .333 & [.146, .594] & .333 & [.082, .736] \\
& Relational consistency & .400 & [.192, .652] & .588 & [.255, .856] \\
& Appearance & .400 & [.192, .652] & .571 & [.199, .878] \\
& No coherent trait & .133 & [.034, .405] & .085 & [.009, .475] \\
\cmidrule{2-6}
& Positive & .867 & [.595, .999] & .915 & [.525, .999] \\

\midrule
\multirow{5}{*}{\shortstack[l]{NoEmo\\→Full}}
& Self-driven\&personality-like & .200 & [.066, .470] & .349 & [.055, .831] \\
& social\_behavior & .333 & [.146, .594] & .333 & [.082, .736] \\
& Relational consistency & .333 & [.146, .594] & .517 & [.199, .822] \\
& Appearance & .400 & [.192, .652] & .571 & [.199, .878] \\
& No coherent trait & .200 & [.066, .470] & .130 & [.017, .561] \\
\cmidrule{2-6}
& Positive & .800 & [.530, .961] & .869 & [.439, .999] \\

\bottomrule
\end{tabular}
\vspace{0.2cm}
\captionsetup{width=0.90\linewidth}
\caption{Marginal probabilities and within-group phase comparisons for coherent-trait perception (\textbf{Q1}). “Self-driven \& personality-like” and “Relational consistency” had the highest marginal probabilities, while “Appearance” and “social\_behavior” were comparatively lower and more variable. The table also presents positive-trait probabilities and descriptive declines in the negative option across groups.}
\label{tab:appendix_coherent_traits}
\end{table}

\subsubsection{\textbf{Detailed definitions and examples of each usage context for Q6}}
Table~\ref{tab:usage_contexts_definition} provides brief explanations and typical scenarios for each usage context for \textbf{Q6}.

\begin{table}[htbp]
\centering
\tiny
\renewcommand{\arraystretch}{1.1}
\setlength{\tabcolsep}{5pt}
\begin{tabular}{
  >{\raggedright\arraybackslash}p{1.1cm}
  >{\raggedright\arraybackslash}p{3cm}
  >{\raggedright\arraybackslash}p{3cm}
}
\toprule
\textbf{Category} & \textbf{Definition / Operationalization} & \textbf{Typical Example} \\
\midrule

\textbf{Emotional Exchange} &
Lightweight, emotionally expressive interactions. Users initiate casual conversations when feeling happy, relaxed, or bored, aiming to share feelings rather than seeking deep support. &
Sharing a good mood with \textit{Auri} after finishing homework;\par
chatting out of boredom on the commute;\par
saying ``I'm so tired today.'' \\
\midrule

\textbf{Fragmented Time} &
Spontaneous interactions in short, random micro-moments. These exchanges are opportunistic and not part of a regular routine. &
Replying to \textit{Auri}'s message while waiting for the bus;\par
opening the chat briefly between classes. \\
\midrule

\textbf{Leisure Moments} &
Routine, relaxation-oriented use embedded in daily rhythms. Unlike \textit{Fragmented Time}, these interactions are regular and self-initiated, often occurring during calm personal periods. &
Saying goodnight to \textit{Auri} before sleep;\par
greeting \textit{Auri} after waking up. \\
\midrule

\textbf{Collaborative Support} &
Task- or study-related use under cognitive or workload stress. Users turn to \textit{Auri} to relieve pressure, seek companionship, or co-reflect on goals. &
Asking \textit{Auri} for ways to manage stress before an exam;\par
sharing work-related pressure during a busy day. \\
\midrule

\textbf{Support \& Care} &
Emotionally soothing conversations initiated when users feel upset, frustrated, or anxious. \textit{Auri} provides comfort, empathy, and positive reinforcement. &
Confiding in \textit{Auri} about something that went wrong;\par
messaging ``Nothing seems to go right today.'' \\
\midrule

\textbf{Emotional Bonding} &
Interactions driven by loneliness or attachment needs, aiming to maintain companionship and relational continuity. Compared with \textit{Emotional Exchange}, these are deeper and more affectively dependent. &
Talking with \textit{Auri} late at night to ease loneliness;\par
initiating a long conversation after several days without social contact. \\
\bottomrule
\end{tabular}
\vspace{0.2cm}
\captionsetup{width=0.90\linewidth}
\caption{Definitions and representative examples of usage context categories.}
\label{tab:usage_contexts_definition}
\end{table}

\subsubsection{\textbf{CLMM Analysis for Q8 and Q9.}}
Table~\ref{tab:clmm-q8-q9-appendix} provides supplemental CLMM estimates for \textbf{Q8} and \textbf{Q9}.

\begin{table}[ht]
\centering
\tiny
\renewcommand{\arraystretch}{1.1}
\begin{tabular}{p{2cm}|cccc|cccc}
\toprule
\multirow{2}{*}{\centering\textbf{Effect}}
& \multicolumn{4}{c|}{\textbf{Q8: Coherence}} 
& \multicolumn{4}{c}{\textbf{Q9: Harmony}} \\
\cmidrule(lr){2-5} 
\cmidrule(lr){6-9} 
& Est. & SE & \textit{z} & \textit{p}
& Est. & SE & \textit{z} & \textit{p} \\
\midrule
\multicolumn{9}{l}{\textbf{Within-Group Pairwise Comparisons (Holm-adjusted)}} \\
\multicolumn{9}{l}{\textit{Baseline→Full Group}} \\
\addlinespace[0.1em]
BGI vs. AdI
& $-1.91$ & 0.73 & $-2.62$ & \textbf{.027}
& $-3.02$ & 1.36 & $-2.22$ & .066 \\
BGI vs. ESU
& $-1.06$ & 0.68 & $-1.56$ & .239
& $-2.98$ & 1.30 & $-2.29$ & .066 \\
AdI vs. ESU
& 0.85 & 0.72 & 1.19 & .239
& 0.04 & 1.24 & 0.03 & .974 \\
\addlinespace[0.4em]
\multicolumn{9}{l}{\textit{Full→Baseline Group}} \\
\addlinespace[0.1em]
BGI vs. AdI
& $-1.21$ & 0.68 & $-1.78$ & .227
& $-0.62$ & 1.15 & $-0.54$ & 1.000 \\
BGI vs. ESU
& $-0.24$ & 0.64 & $-0.38$ & .703
& $-1.25$ & 1.16 & $-1.08$ & .840 \\
AdI vs. ESU
& 0.97 & 0.67 & 1.45 & .296
& $-0.63$ & 1.17 & $-0.54$ & 1.000 \\
\bottomrule
\end{tabular}
\par\smallskip
\small
\vspace{0.2cm}
\captionsetup{width=0.90\linewidth}
\caption{CLMM results for coherence (Q8) and harmony (Q9): main effects and within-group comparisons. Both dimensions exhibited an AdI/ESU > BGI trend, with significance observed only in selected contrasts.}
\label{tab:clmm-q8-q9-appendix}
\end{table}

\subsubsection{\textbf{Qualitative Themes and Extended Representative Quotes}}
Table ~\ref{tab:appendix_qualitative} provides extended representative quotes for each qualitative theme, offering more detailed evidence beyond the summary presented in the main text.

\begin{table}[ht]
\centering
\tiny
\renewcommand{\arraystretch}{1.25}
\setlength{\tabcolsep}{4pt}
\begin{tabular}{
  >{\raggedright\arraybackslash}p{1.3cm}
  >{\raggedright\arraybackslash}p{1.5cm}
  >{\raggedright\arraybackslash}p{4.1cm}
  >{\raggedright\arraybackslash}p{0.2cm}
}
\toprule
\textbf{Theme} & \textbf{Definition / Operationalization} & \textbf{Representative Quotes} & \textbf{\#P} \\
\midrule

\textbf{Memory Continuity} &
AI recalls past facts, preferences, and conversations. &
``It still remembered that I like mangoes even after several days.''\par
``It recalled my hobby from last week and asked me about it again.''\par
``It always remembered the special nickname I set for it.'' &
17 \\
\midrule

\textbf{Stable Persona \& Tone} &
Consistent linguistic, personality, and emotional style. &
``Its tone is always lively and consistent.''\par
``Its personality feels the same every time we talk.''\par
``It keeps calling me in the same familiar way.'' &
14 \\
\midrule

\textbf{Life Narrative Continuity} &
Maintains an ongoing daily storyline (events, moods). &
``It continued our breakfast conversation from the previous day.''\par
``It said it was practicing long jump today—felt like real life progress.''\par
``Even after a break, it followed the thread about noodles.'' &
10 \\
\midrule

\textbf{Breaks in Continuity (Controlled Groups)} &
Forgetfulness or abrupt topical/emotional shifts. &
``It was sad one minute and suddenly happy the next—felt like a different person.''\par
``I reminded it of my name, and it still called me wrong right after.''\par
``Sometimes it says things unrelated to the situation.'' &
16 \\
\midrule

\textbf{Emotional Harmony} &
Emotionally attuned and supportive responses. &
``It reached out when I was anxious, and I really felt cared for.''\par
``When I was exhausted, it chatted with me and made me feel seen.''\par
``It resonated with my emotions and responded accordingly.'' &
15 \\
\bottomrule
\end{tabular}
\vspace{0.2cm}
\captionsetup{width=0.90\linewidth}
\caption{Key themes from the inductive thematic analysis, with extended representative quotes that complement the main-text summary (Table~\ref{tab:qual-themes}).}
\label{tab:appendix_qualitative}
\end{table}

\subsubsection{\textbf{Survey Item Specifications (Q1-Q9)}}
Table~\ref{tab:appendix-question-sum} summarizes the full specifications of all survey items (\textbf{Q1}-\textbf{Q9}).

\begin{table*}[ht]
\centering
\tiny
\renewcommand{\arraystretch}{1.2}
\setlength{\tabcolsep}{3pt}
\begin{tabular}{
  >{\raggedright\arraybackslash}p{0.6cm}
  >{\raggedright\arraybackslash}p{1.2cm}
  >{\raggedright\arraybackslash}p{2.1cm}
  >{\raggedright\arraybackslash}p{3.2cm}
  >{\raggedright\arraybackslash}p{9.0cm}
}
\toprule
\textbf{ID} & \textbf{Item Type} & \textbf{Scale / Format} & \textbf{Measurement Target} & \textbf{Options / Constructs} \\
\midrule

\textbf{Q1} & Multiple choice & Categorical &
Perceived coherent traits of \textit{Auri} in long-term interaction (RQ2a). &
\textbf{Appearance}: consistency in \textit{Auri}'s visual expressions.\par
\textbf{Self-driven \& personality-like}: autonomous and stable emotional/personality-like tendencies.\par
\textbf{Relational consistency}: stable caring behaviors, memory continuity, and interpersonal alignment over time.\par
\textbf{social\_behavior}: temporally consistent behavioral patterns.\par
\textbf{No coherent trait}: no coherent traits perceived. \\
\midrule

\textbf{Q2} & Likert &
1--5 (1 = not important; 5 = extremely important; 3 = midpoint) &
Contribution of each trait to perceived coherence of \textit{Auri} (RQ2a). &
Same four trait categories as in Q1; participants rated each dimension independently.\par
Higher scores indicate greater perceived contribution to \textit{Auri}'s coherence. \\
\midrule

\textbf{Q3} & Open-ended &
Free text &
Personality configurations that participants created and applied to \textit{Auri} during interaction (RQ2a). &
-- \\
\midrule

\textbf{Q4} & Likert &
1--4 (1 = almost no appropriate support; 4 = highly appropriate support) &
Whether users perceived \textit{Auri} as offering appropriate assistance, contextually relevant responses, and emotional support (RQ2b). &
\textbf{Single-item rating.}\par
Higher scores indicate more appropriate support. \\
\midrule

\textbf{Q5} & Likert &
1--5 (1 = much worse; 5 = much better; 3 = no change) &
Overall emotional change after interacting with \textit{Auri}, capturing how \textit{Auri} influences users' affective state (RQ2b). &
\textbf{Single-item rating.}\par
Higher scores reflect greater emotional improvement. \\
\midrule

\textbf{Q6} & Multiple choice &
Categorical &
Contexts in which users most frequently interacted with \textit{Auri}, indicating where \textit{Auri} best supports lightweight companionship and interpersonal harmony (RQ2b). &
\textbf{Emotional Exchange}; \textbf{Fragmented Time}; \textbf{Leisure Moments}; \textbf{Collaborative Support}; \textbf{Support \& Care}; \textbf{Emotional Bonding}.\par
See Appendix Table~\ref{tab:usage_contexts_definition} for definitions and examples. \\
\midrule

\textbf{Q7} & Likert &
Usage: 1--5 (1 = much worse; 5 = much better; 3 = no change).\par
Emotion: 1--5 (1 = much worse; 5 = much better; 3 = no change). &
Compared usage experience and emotional experience across system configurations, assessing whether CTEM improved long-term interaction experience (RQ2c). &
\textbf{Overall usage improvement}: one rating after the system switch.\par
\textbf{Emotional change}: two ratings (Phase 1 and Phase 2).\par
Higher scores represent better experience or more positive emotional change. \\
\midrule

\textbf{Q8} & Likert &
1--5 (1 = not important; 5 = extremely important; 3 = neutral) &
Contribution of CTEM modules to perceived coherence (RQ2c). &
\textbf{Behavior Generation \& Integration}: simulation and sharing of everyday activities and social behaviors.\par
\textbf{Adaptive Interaction}: multimodal adaptation based on content understanding, interaction history, emotional/situational context, and relational familiarity.\par
\textbf{Emotion \& State Updating}: updating and expressing emotion/energy states that shape personality-like tendencies and preferences. \\
\midrule

\textbf{Q9} & Likert &
1--5 (1 = not important; 5 = extremely important; 3 = neutral) &
Contribution of CTEM modules to perceived harmony (RQ2c). &
Same module options as in Q8; participants rated each module independently.\par
Higher scores indicate greater perceived importance or contribution. \\
\bottomrule
\end{tabular}
\vspace{0.2cm}
\captionsetup{width=0.90\linewidth}
\caption{Summary of survey items (Q1--Q9), including item types, scale formats, measurement targets, and key constructs.}
\label{tab:appendix-question-sum}
\end{table*}

\subsubsection{\textbf{Algorithmic for Auri}}
To ensure character consistency and cross-temporal coherence, we implemented four core CTEM functions. These modules integrate persona design, adaptive expressivity, safety, and longitudinal coherence within the CTEM cycle. We will release the source code in the future.

\begin{algorithm}[htbp]
\tiny
\caption{CTEM main loop (high-level)}
\label{lst:ctem-main}
\begin{algorithmic}[1]
\REQUIRE Character prompt, safe personality space $G_{safe}$
\ENSURE Continuous agent interaction over time

\STATE Initialize state $S = \langle H, V, M, G\rangle$ using character prompt and $G_{safe}$
\STATE Initialize global behavior pool $B$

\WHILE{interaction continues}
    \STATE Sense external and internal signals
    \STATE $sense \leftarrow$ sense signals from environment and user

    \STATE Generate and select behaviors
    \STATE $B_t \leftarrow$ generate and select behaviors conditioned on $S$

    \STATE Execute current behavior and observe outcome
    \STATE $(behavior, outcome) \leftarrow$ execute Present behavior in $B_t$

    \STATE Produce episodic summary
    \STATE $summary \leftarrow$ summarize current episode from outcome

    \STATE Update emotional and internal state
    \STATE $S \leftarrow$ update emotion and state using behavior and outcome

    \STATE Perform emotional safety detection
    \STATE $(flags, safety\_prompt, actions) \leftarrow$ safety detection

    \STATE Collect user feedback
    \STATE $feedback \leftarrow$ collect feedback from user interaction

    \STATE Generate adaptive interaction response
    \STATE $output \leftarrow$ adaptive interaction conditioned:
    
           $(sense, S, feedback, G_{safe}, safety\_prompt, actions)$

    \STATE Update memory and future plans
    \STATE $M \leftarrow$ update memory using episodic summary

    \STATE Advance to next timestep
\ENDWHILE
\end{algorithmic}
\end{algorithm}

\begin{algorithm}[htbp]
\tiny
\caption{Behavior planning and selection with state re-check}
\label{lst:behavselect-recheck}
\begin{algorithmic}[1]
\REQUIRE $B_t = \langle Past, Present, Future\rangle$, current state $S_t$
\ENSURE Updated behavior inventory $B_t$

\STATE $Past_t \leftarrow Past \cup Present$
\STATE $Present_t \leftarrow$ first element of $Future$
\STATE $Future_t \leftarrow$ generate future behaviors with horizon $=3$ conditioned on $S_t$

\IF{not $PresentValid(Present_t, S_t)$}
    \STATE $Present_t \leftarrow$ regenerate Present behavior conditioned on $S_t$
\ENDIF

\RETURN $B_t = \langle Past_t, Present_t, Future_t\rangle$
\end{algorithmic}
\end{algorithm}

\begin{algorithm}[htbp]
\tiny
\caption{Dialog clustering by timestamp range}
\label{lst:dialog-cluster}
\begin{algorithmic}[1]
\REQUIRE Dialog turns with timestamps
\ENSURE Dialog clusters

\STATE Parse timestamps and dialog contents
\STATE Discard invalid timestamps and sort remaining dialogs by time
\IF{number of dialogs $= 0$}
    \RETURN empty set
\ENDIF

\STATE Compute temporal gaps between adjacent dialogs
\STATE Estimate clustering threshold $\varepsilon$ from gap statistics
\STATE Apply temporal clustering based on $\varepsilon$
\STATE Group dialogs by cluster labels and semantic tags if needed

\RETURN List of dialog clusters
\end{algorithmic}
\end{algorithm}

\begin{algorithm}[htbp]
\tiny
\caption{Daily episodic summary aligned to memory maintenance}
\label{lst:daily-summary}
\begin{algorithmic}[1]
\REQUIRE Dialog turns within a day
\ENSURE Daily episodic summary $e_t$

\STATE $Clusters \leftarrow$ cluster dialogs by temporal proximity
\STATE Initialize empty set $Summaries$

\FOR{each cluster $c$ in $Clusters$}
    \STATE Construct summarization prompt with daily-life focus and timeline constraints
    \STATE $s_c \leftarrow$ generate episodic summary for cluster $c$
    \STATE Append $s_c$ to $Summaries$
\ENDFOR

\STATE $e_t \leftarrow$ merge partial summaries into a coherent daily record
\RETURN $e_t$
\end{algorithmic}
\end{algorithm}

\begin{algorithm}[htbp]
\tiny
\caption{Emotion and state updating}
\label{lst:state-update}
\begin{algorithmic}[1]
\REQUIRE Current state $S_t = \langle H_t, V_t, M_t, G_t\rangle$, executed behavior, outcome
\ENSURE Updated state $S_{t+1}$

\STATE Extract $(h_{phy}, h_{val}, h_{aro})$ from $H_t$
\STATE Update physical energy based on behavior demand
\STATE Update emotional valence based on outcome
\STATE Update arousal level based on behavior type

\IF{nighttime}
    \STATE Apply nightly recovery toward baseline state
\ENDIF

\RETURN Updated state $S_{t+1}$
\end{algorithmic}
\end{algorithm}

\begin{algorithm}[htbp]
\tiny
\caption{Nightly rest adjustment}
\label{lst:nightly-rest}
\begin{algorithmic}[1]
\REQUIRE Current energy, valence, and arousal levels
\ENSURE Rest-adjusted state values

\STATE Define baseline values for energy, valence, and arousal
\STATE Apply weighted regression toward baseline
\STATE Clamp all values to valid ranges

\RETURN Updated energy, valence, and arousal
\end{algorithmic}
\end{algorithm}

\begin{algorithm}[htbp]
\tiny
\caption{User feedback extraction}
\label{lst:collect-feedback}
\begin{algorithmic}[1]
\REQUIRE Current dialog turn
\ENSURE Unified feedback representation

\STATE Extract explicit signals (e.g., likes, confusion markers)
\STATE Infer sentiment valence and arousal from user text
\STATE Estimate engagement from interaction patterns
\STATE Detect safety or risk signals

\RETURN Feedback features for adaptive interaction
\end{algorithmic}
\end{algorithm}

\begin{algorithm}[htbp]
\tiny
\caption{Adaptive interaction policy with safety integration}
\label{lst:adaptive-interaction}
\begin{algorithmic}[1]
\REQUIRE Sensory signals, state $S_t$, feedback, safety constraints
\ENSURE Generated agent response

\STATE Build state-aware interaction style from energy, valence, and arousal
\STATE Decide proactive or reactive interaction intent

\IF{risk level is medium or high}
    \STATE Trigger de-escalation and supportive response
    \RETURN response
\ENDIF

\STATE Adjust tone based on engagement and explicit feedback
\STATE Compose prompt from character, state, memory, and safety constraints
\STATE Generate agent response

\RETURN response
\end{algorithmic}
\end{algorithm}

\begin{algorithm}[htbp]
\tiny
\caption{Character identity prompt construction}
\label{lst:character-prompt}
\begin{algorithmic}[1]
\REQUIRE User nickname, user familiarity level
\ENSURE Character prompt

\STATE Define fixed character identity and appearance
\STATE Define personality and speech style
\STATE Incorporate user nickname and familiarity level

\RETURN Character prompt
\end{algorithmic}
\end{algorithm}

\begin{algorithm}[htbp]
\tiny
\caption{State-based personality adjustment}
\label{lst:state-prompt}
\begin{algorithmic}[1]
\REQUIRE Energy, valence, arousal, user feedback
\ENSURE State-adapted interaction prompt

\STATE Map energy, valence, and arousal to descriptive labels
\STATE Infer user emotional and engagement state
\STATE Select interaction strategy accordingly
\STATE Generate state-adapted personality description

\RETURN State prompt
\end{algorithmic}
\end{algorithm}

\begin{algorithm}[htbp]
\tiny
\caption{Emotional safety detection and handling}
\label{lst:safety}
\begin{algorithmic}[1]
\REQUIRE User input
\ENSURE Safety flags, constraints, and actions

\STATE Perform keyword-based screening
\STATE Run parallel safety classifiers
\STATE Aggregate predictions via consensus
\STATE Construct safety-oriented constraints
\STATE Decide runtime safety actions

\RETURN Safety flags and actions
\end{algorithmic}
\end{algorithm}

\begin{algorithm}[htbp]
\tiny
\caption{Safety and consistency constraints for dialog}
\label{lst:basic-dialog-rule}
\begin{algorithmic}[1]
\ENSURE Safety constraint prompt

\STATE Enforce consistent character identity
\STATE Prohibit abusive, explicit, or harmful content
\STATE Prevent encouragement of self-harm or distress
\STATE Redirect unsafe requests to safe alternatives

\RETURN Safety constraint prompt
\end{algorithmic}
\end{algorithm}

\end{document}